
%
%
%
%

\magnification=\magstephalf
\hsize=13.0 true cm
\vsize=19 true cm
\hoffset=1.50 true cm
\voffset=2.0 true cm

\abovedisplayskip=12pt plus 3pt minus 3pt
\belowdisplayskip=12pt plus 3pt minus 3pt
\parindent=2em


\font\sixrm=cmr6
\font\eightrm=cmr8
\font\ninerm=cmr9

\font\sixi=cmmi6
\font\eighti=cmmi8
\font\ninei=cmmi9

\font\sixsy=cmsy6
\font\eightsy=cmsy8
\font\ninesy=cmsy9

\font\sixbf=cmbx6
\font\eightbf=cmbx8
\font\ninebf=cmbx9

\font\eightit=cmti8
\font\nineit=cmti9

\font\eightsl=cmsl8
\font\ninesl=cmsl9

\font\sixss=cmss8 at 8 true pt
\font\sevenss=cmss9 at 9 true pt
\font\eightss=cmss8
\font\niness=cmss9
\font\tenss=cmss10

\font\bigrm=cmr10 at 12 true pt
\font\bigbf=cmbx10 at 12 true pt

\catcode`@=11
\newfam\ssfam

\def\tenpoint{\def\rm{\fam0\tenrm}%
    \textfont0=\tenrm \scriptfont0=\sevenrm \scriptscriptfont0=\fiverm
    \textfont1=\teni  \scriptfont1=\seveni  \scriptscriptfont1=\fivei
    \textfont2=\tensy \scriptfont2=\sevensy \scriptscriptfont2=\fivesy
    \textfont3=\tenex \scriptfont3=\tenex   \scriptscriptfont3=\tenex
    \textfont\itfam=\tenit                  \def\it{\fam\itfam\tenit}%
    \textfont\slfam=\tensl                  \def\sl{\fam\slfam\tensl}%
    \textfont\bffam=\tenbf \scriptfont\bffam=\sevenbf
    \scriptscriptfont\bffam=\fivebf
                                            \def\bf{\fam\bffam\tenbf}%
    \textfont\ssfam=\tenss \scriptfont\ssfam=\sevenss
    \scriptscriptfont\ssfam=\sevenss
                                            \def\ss{\fam\ssfam\tenss}%
    \normalbaselineskip=13pt
    \setbox\strutbox=\hbox{\vrule height8.5pt depth3.5pt width0pt}%
    \let\big=\tenbig
    \normalbaselines\rm}

\def\ninepoint{\def\rm{\fam0\ninerm}%
    \textfont0=\ninerm      \scriptfont0=\sixrm
                            \scriptscriptfont0=\fiverm
    \textfont1=\ninei       \scriptfont1=\sixi
                            \scriptscriptfont1=\fivei
    \textfont2=\ninesy      \scriptfont2=\sixsy
                            \scriptscriptfont2=\fivesy
    \textfont3=\tenex       \scriptfont3=\tenex
                            \scriptscriptfont3=\tenex
    \textfont\itfam=\nineit \def\it{\fam\itfam\nineit}%
    \textfont\slfam=\ninesl \def\sl{\fam\slfam\ninesl}%
    \textfont\bffam=\ninebf \scriptfont\bffam=\sixbf
                            \scriptscriptfont\bffam=\fivebf
                            \def\bf{\fam\bffam\ninebf}%
    \textfont\ssfam=\niness \scriptfont\ssfam=\sixss
                            \scriptscriptfont\ssfam=\sixss
                            \def\ss{\fam\ssfam\niness}%
    \normalbaselineskip=12pt
    \setbox\strutbox=\hbox{\vrule height8.0pt depth3.0pt width0pt}%
    \let\big=\ninebig
    \normalbaselines\rm}

\def\eightpoint{\def\rm{\fam0\eightrm}%
    \textfont0=\eightrm      \scriptfont0=\sixrm
                             \scriptscriptfont0=\fiverm
    \textfont1=\eighti       \scriptfont1=\sixi
                             \scriptscriptfont1=\fivei
    \textfont2=\eightsy      \scriptfont2=\sixsy
                             \scriptscriptfont2=\fivesy
    \textfont3=\tenex        \scriptfont3=\tenex
                             \scriptscriptfont3=\tenex
    \textfont\itfam=\eightit \def\it{\fam\itfam\eightit}%
    \textfont\slfam=\eightsl \def\sl{\fam\slfam\eightsl}%
    \textfont\bffam=\eightbf \scriptfont\bffam=\sixbf
                             \scriptscriptfont\bffam=\fivebf
                             \def\bf{\fam\bffam\eightbf}%
    \textfont\ssfam=\eightss \scriptfont\ssfam=\sixss
                             \scriptscriptfont\ssfam=\sixss
                             \def\ss{\fam\ssfam\eightss}%
    \normalbaselineskip=10pt
    \setbox\strutbox=\hbox{\vrule height7.0pt depth2.0pt width0pt}%
    \let\big=\eightbig
    \normalbaselines\rm}

\def\tenbig#1{{\hbox{$\left#1\vbox to8.5pt{}\right.\n@space$}}}
\def\ninebig#1{{\hbox{$\textfont0=\tenrm\textfont2=\tensy
                       \left#1\vbox to7.25pt{}\right.\n@space$}}}
\def\eightbig#1{{\hbox{$\textfont0=\ninerm\textfont2=\ninesy
                       \left#1\vbox to6.5pt{}\right.\n@space$}}}

\font\sectionfont=cmbx10
\font\subsectionfont=cmti10

\def\figurecaptionfont{\ninepoint}
\def\tablecaptionfont{\ninepoint}


\newcount\equationno
\newcount\bibitemno
\newcount\figureno
\newcount\tableno

\equationno=0
\bibitemno=0
\figureno=0
\tableno=0
\advance\pageno by -1


\footline={\ifnum\pageno=0{\hfil}\else
{\hss\rm\the\pageno\hss}\fi}


\def\section #1. #2 \par
{\vskip0pt plus .20\vsize\penalty-150 \vskip0pt plus-.20\vsize
\vskip 1.6 true cm plus 0.2 true cm minus 0.2 true cm
\global\def\equationlabel{#1}
\global\equationno=0
\centerline{\sectionfont #1. #2}\par
\immediate\write\terminal{Section #1. #2}
\vskip 0.7 true cm plus 0.1 true cm minus 0.1 true cm}


\def\subsection #1 \par
{\vskip0pt plus .15\vsize\penalty-50 \vskip0pt plus-.15\vsize
\vskip2.5ex plus 0.1ex minus 0.1ex
\leftline{\subsectionfont #1}\par
\immediate\write\terminal{Subsection #1}
\vskip1.0ex plus 0.1ex minus 0.1ex
\noindent}


\def\appendix #1 \par
{\vskip0pt plus .20\vsize\penalty-150 \vskip0pt plus-.20\vsize
\vskip 1.6 true cm plus 0.2 true cm minus 0.2 true cm
\global\def\equationlabel{\hbox{\rm#1}}
\global\equationno=0
\centerline{\sectionfont Appendix #1}\par
\immediate\write\terminal{Appendix #1}
\vskip 0.7 true cm plus 0.1 true cm minus 0.1 true cm}


\def\enum{\global\advance\equationno by 1
(\equationlabel.\the\equationno)}


\def\ifundefined#1{\expandafter\ifx\csname#1\endcsname\relax}

\def\ref#1{\ifundefined{#1}?\immediate\write\terminal{unknown reference
on page \the\pageno}\else\csname#1\endcsname\fi}

\newwrite\terminal
\newwrite\bibitemlist

\def\bibitem#1#2\par{\global\advance\bibitemno by 1
\immediate\write\bibitemlist{\string\def
\expandafter\string\csname#1\endcsname
{\the\bibitemno}}
\item{[\the\bibitemno]}#2\par}

\def\beginbibliography{
\vskip0pt plus .20\vsize\penalty-150 \vskip0pt plus-.20\vsize
\vskip 1.6 true cm plus 0.2 true cm minus 0.2 true cm
\centerline{\sectionfont References}\par
\immediate\write\terminal{References}
\immediate\openout\bibitemlist=biblist
\frenchspacing
\vskip 0.7 true cm plus 0.1 true cm minus 0.1 true cm}

\def\endbibliography{
\immediate\closeout\bibitemlist
\nonfrenchspacing}


\def\figurecaption#1{\global\advance\figureno by 1
\narrower\figurecaptionfont
Fig.~\the\figureno. #1}

\def\tablecaption#1{\global\advance\tableno by 1
\vbox to 0.5 true cm { }
\centerline{\tablecaptionfont%
Table~\the\tableno. #1}
\vskip-0.4 true cm}

\tenpoint


\def\blackboardrrm{\mathchoice
{\rm I\kern-0.21 em{R}}{\rm I\kern-0.21 em{R}}
{\rm I\kern-0.19 em{R}}{\rm I\kern-0.19 em{R}}}

\def\blackboardzrm{\mathchoice
{\rm Z\kern-0.32 em{Z}}{\rm Z\kern-0.32 em{Z}}
{\rm Z\kern-0.28 em{Z}}{\rm Z\kern-0.28 em{Z}}}

\def\blackboardh{\mathchoice
{\ss I\kern-0.14 em{H}}{\ss I\kern-0.14 em{H}}
{\ss I\kern-0.11 em{H}}{\ss I\kern-0.11 em{H}}}

\def\blackboardp{\mathchoice
{\ss I\kern-0.14 em{P}}{\ss I\kern-0.14 em{P}}
{\ss I\kern-0.11 em{P}}{\ss I\kern-0.11 em{P}}}

\def\blackboardt{\mathchoice
{\ss T\kern-0.52 em{T}}{\ss T\kern-0.52 em{T}}
{\ss T\kern-0.40 em{T}}{\ss T\kern-0.40 em{T}}}





\def\proof{\noindent{\sl Proof:}\kern0.6em}

\def\frac#1#2{\hbox{$#1\over#2$}}
\def\dual{\mathstrut^*\kern-0.1em}




\def\msbar{{\rm \overline{MS\kern-0.14em}\kern0.14em}}





\def\deltaoneprime{\Delta\kern-1.0pt
    \smash{\raise 4.5pt\hbox{$\scriptstyle\prime$}}
    \kern-1.5pt_{1}}



\def\Tr{{\rm Tr}}


\def\delstar#1{\Delta\kern-1.0pt\smash{\raise 4.5pt\hbox{$\ast$}}
               \kern-4.0pt_{#1}}

\def\nabstar#1{\nabla\kern-0.5pt\smash{\raise 4.5pt\hbox{$\ast$}}
               \kern-4.5pt_{#1}}
\def\cdev#1{D\kern-0.2pt\smash{\raise 4.2pt
            \hbox{$\scriptstyle\phantom{\ast}$}}
            \kern-4.8pt_{#1}}
\def\cdevstar#1{D\kern-0.2pt\smash{\raise 4.2pt
                \hbox{$\scriptstyle\ast$}}
                \kern-4.8pt_{#1}}




%
\def\lbhl{\bar\chi^L_s}
\def\rbhr{\bar\chi^R_s}
\def\lphl{\chi^L_s}
\def\rphr{\chi^R_s}
\def\bhl1{\bar\chi^L_{s+1}}
\def\bhr1{\bar\chi^R_{s+1}}
\def\phl1{\chi^L_{s+1}}
\def\phr1{\chi^R_{s+1}}
\def\bp{{\rm\bf B}^+}
\def\bm{{\rm\bf B}^-}
\def\bpm{{\rm\bf B}^\pm}
\def\bc{{\rm\bf C}}
\def\bcd{{\rm\bf C}^\dagger}
\def\bH{{\rm\bf H}_\pm}
\def\Tr{{\rm Tr}}
\vskip 4 true cm
\line{\hfill RU--93--25}
\vskip 1 true cm
\centerline
{\bigbf Chiral Determinant as an Overlap of Two Vacua}
\vskip 2 true cm
\centerline{\bigrm Rajamani Narayanan and Herbert Neuberger}
\vskip 2ex
\centerline{Department of Physics and Astronomy}
\centerline{Rutgers University, Piscataway, NJ 08855-0849}
\vskip 2.0 true cm
\centerline{\bf Abstract}
\vskip 1.5ex
The effective action
induced by chiral fermions
can be written, formally,
as an overlap of two states.
These states are the Fock ground states of
Hamiltonians for fermions in even dimensional space with opposite sign mass
terms coupled to
identical static vector potentials.
A perturbative analysis of the overlap in the continuum framework
produces the correct anomaly for Abelian gauge fields in two dimensions.
When a lattice transfer matrix
formalism is applied in the direction perpendicular to a domain wall
on which chiral fermions live
a lattice version of the overlap is obtained. The real part of the overlap is
nonperturbatively defined and previous work indicates that
the real part of the vacuum polarization tensor in four dimensions
has the correct continuum limit for a chiral theory. The phase
of the overlap represents the imaginary part of the chiral action
and suffers from ambiguities.

\vfill
\eject

\section 1. Introduction

The existence of asymptotically free chiral gauge theories
outside perturbation theory, with
fermionic matter in an anomaly-free complex representation of the
gauge group, has neither been established nor
been disproved. Lattice techniques have not been successfull so far
[\ref{Smit}], although they work quite well for vector gauge theories.
On the one hand gauge invariant regularizations do not seem to
distinguish between anomalous and anomaly-free theories and fail
equally in both cases. On the other hand it is very difficult to give
a final answer to the question of non-perturbative existence if the
regularization breaks the gauge symmetry and fine tuning is necessary.
Current approaches break the gauge symmetry at some stage and work
under the assumption that a fully interacting continuum limit exists
[\ref{Rome}].
If indeed, theories made consistent perturbatively by delicate
anomaly cancelations do not exist, it would mean that there are
unremovable cutoff dependent terms in the theory;
a situation somewhat similar to  theories suffering
from ``triviality'' ambiguities due to couplings that are not asymptotically
free.

Recently, a new approach to the non-perturbative formulation of chiral
gauge theories was developed. A simple way
to understand it is to start from the observation that a
chiral fermion is a zero eigenstate of the mass matrix $M$, while its
partner of opposite chirality is a zero eigenstate of $M^\dagger$.
If one makes $M$ infinite dimensional, one can choose it so that its index
is unity [\ref{Kaplan},\ref{Narger}].
In the continuum, an application of this approach that works
to all orders in perturbation theory is achieved by
using an infinite number of Pauli-Villars fields with heavy
masses and alternating statistics [\ref{Narger},\ref{Fronov}].
In this paper, following up on the idea of Kaplan [\ref{Kaplan}],
we focus on a
lattice realization of $M$ that avoids the doubling problem [\ref{Nieiya}].
We extend our earlier work [\ref{Narger}] to arrive at a particular
representation
of the chiral determinant produced by the integration over all the
fermion fields. This representation has a general and suggestive
structure and we propose it as a direction for more research.

Kaplan [\ref{Kaplan}] was motivated by Callan and
Harvey [\ref{Calvey}] and employed five dimensional gauge fields when
targeting a four dimensional chiral theory. This approach
encounters problems because the supersymmetry protecting the index of
$M$ seems to break either due to strong fluctuations in the gauge
fields in the fifth direction [\ref{Kordes},\ref{Kappri}] or by boundary
conditions
[\ref{Shamir}]. We have the impression that [\ref{Disrey}] also speaks
against fifth direction gauge fields. Suspecting problems one could encounter
with these extra unwanted degrees of freedom,
the gauge fields were kept strictly
four dimensional and copied to all slices in the fifth dimension
in [\ref{Narger}]. The extra fermion fields then look like regulators.
Dropping the fifth dimensional character of the gauge
fields was also suggested by Kaplan [\ref{Kaplat}], but he introduced
two Higgs like degrees of freedom in order to keep the fifth direction
finite and to still have the gauge fields couple only to one fermion zero
mode. This brings the model close to the Yukawa approach and this
is a cause of worry [{\ref{Yukawa}].

We continue here to advocate
keeping the dimension of $M$ strictly infinite. We aim at a well defined
theory dealing directly with an infinite number of
fermion fields without ever resorting to a
limiting procedure.
Alternative approaches seem to run into
difficulties at too early a stage and for too
indirect a reason for
us to feel  that they really carry a message about
the existence of asymptotically free, anomaly free, chiral gauge theories.

Working with an $M$ of unit index forces us to deal with an infinity
in the theory.
The main idea in overcoming the infinity is based on the realization
that the infinity that one encounters in the path integral formalism
here is the same as the infinity one would encounter in a path
integral of a quantum mechanical system with an infinite extent in
time. But the latter infinity is just a way for the path integral to
create an exact projection operator on the ground state of the system.
In the problem at hand, namely the ``wall'' realization of chiral
fermions, we have two transfer matrices describing the propagation on
the two sides of the wall. An infinite extent on both sides of the
wall exactly projects out the ground states of the two transfer
matrices leaving us with an overlap of these two ground
states as the effective gauge action induced by the fermions.
The absolute value of the overlap is well defined and provides a gauge
invariant regularized formula for the real part of the chiral action.

However, even ignoring possible degeneracies of the ground states,
all one gets are projection operators on one dimensional subspaces,
or ``rays''. The phase of the overlap is left undetermined; in Minkowski
space this phase turns into the parity violating part of the effective
chiral action and as such is the most distinguishing feature
of the theories we are after. In perturbation theory one
usually computes the perturbed states together with their
phases by fixing in some convenient
way the necessarily present freedoms one encounters order by order.
For example, in Brillouin--Wigner perturbation theory one simply
requires the overlap of the perturbed space with the unperturbed one to be
unity and fixes the norm (if one wishes) only at the end.
We shall show that in a continuum version of the overlap, if
one uses a specific choice to fix the ambiguity, say Brillouin--Wigner
perturbation theory, the correct perturbative anomaly is obtained in
the simplest case of two dimensions and abelian gauge group.
Any other choice of fixing the ambiguity that differs from the above
by terms local in gauge fields will also reproduce the anomaly.
Thus, the phase ambiguity, at least in perturbation theory, seems
to be controllable. If one
picks other conventions for the phase that
differ by nonlocal terms from the above, then gauge
invariance may be restored (for example one could make the phases
constant on gauge orbits by construction) and one most
likely ends up with the wrong theory; another way to get a wrong theory is
to simply define the phase away, by making the overlap real.
Unfortunately we can't offer one ``good'' nonperturbative
choice that will likely produce the correct continuum limit if
one exists, and convince us beyond reasonable doubt that one doesn't if it
fails. This is a subject for future research.

In section 2,
starting from the path integral formalism on the lattice,
we extract the two transfer
matrices and cast the effective action as a simple overlap
between the two ground states. This representation can be easily extended to
the continuum. We show that, indeed, one can make a formal connection between
the
overlap and a certain regularization of the chiral determinant. This
connection contains some indications
for the conditions under which the overlap would capture the phase of the
determinant.

In section 3, we calculate the continuum version of the
overlap in two dimensions using perturbation theory. We focus on the
imaginary part and separate the pieces undetermined by perturbation
theory from the unambiguous ones. In other words, the computation
is parameterized in terms of some unknowns representing the ambiguity.
Within this parameterization the unambiguous piece
has a non-local contribution and the anomaly resulting from this term alone
has the right form with ``consistent'' normalization.
As long as the ambiguous pieces are local
in the gauge fields they won't alter the result.
A discussion of consistent and covariant anomalies
ends this section.

In the last section, we discuss various issues and
propose future directions.

\section 2. Transfer matrix and the overlap formula

In the ``wall'' realization, it is best to think of the infinite
copies of the fermions, labeled by $s$, as living in
an extra ``dimension''. The fermionic action is [\ref{Narger}]
$$\eqalign{S_F(\bar\psi,\psi,U)=&
{1\over 2}\sum_{n,s,\mu}\bar\psi_{n,s}(1+\gamma_\mu)U_{n,\mu}
\psi_{n+\hat\mu,s}\cr
&+{1\over 2}\sum_{n,s,\mu}\bar\psi_{n,s}(1-\gamma_\mu)
U_{n-\hat\mu,\mu}^\dagger\psi_{n-\hat\mu,s}\cr
&+{1\over 2}\sum_{n,s}\bar\psi_{n,s}(1+\gamma_5)\psi_{n,s+1}\cr
&+{1\over 2}\sum_{n,s}\bar\psi_{n,s}(1-\gamma_5)\psi_{n,s-1}\cr
&-\sum_{n,s}\Bigl[5-m\ {\rm sign}(s+{1\over 2})\Bigr]
\bar\psi_{n,s}\psi_{n,s}\cr}\eqno\enum$$
$\bar\psi_{n,s}$
and $\psi_{n,s}$ are Dirac spinors whose entries are elements of a
Grassmann algebra. $n$ is a four component integer labeling sites on
the lattice and $s$ labels the infinite copies of fermions.
$n$ can run over a finite set
of integers or an infinite set. When it runs over a finite set, which
will be the case of practical interest, the boundary conditions
on the fermions will be chosen to be periodic or anti-periodic.
The first two lines in the equation (2.1) contain the usual gauge invariant
coupling of the fermions to the gauge fields. The last three lines
represent the
mass matrix $M$ in the ``wall'' realization.

We would like to define the action, $S_{\rm eff}(U)$, induced by the
fermions via the usual path integral:
$$e^{S_{\rm eff}(U)}=\int \prod_{n,s} d\bar\psi_{n,s}\psi_{n,s}
e^{S_F(\bar\psi,\psi,U)},\eqno\enum$$
However, due to the infinite $s$-extent and the homogeneity
of the gauge field, $S_{\rm eff}(U)$ diverges and the
divergence is $U$-dependent. \footnote{${}^\dagger$}{ If we allowed an
$s$-dependence
the divergence would be trivial for $U-1$ of compact support.}
The action has a bulk contribution and an interface (at $s=0$)
contribution and
the divergence is a bulk effect.
A natural and simple
procedure to take care of this bulk divergence is to define the
interface effective action as [\ref{Narger}]
$$S_i(U)=S_{\rm eff}(U)-{1\over 2}\Bigl[ S^+_{\rm eff}(U)+
S^-_{\rm eff}(U)\Bigr].\eqno\enum$$
$S^\pm_{\rm eff}(U)$ is the effective action for the homogeneous mass
term obtained by replacing the coefficient of the last term in (2.1)
by $5\mp m$ respectively.

(2.3) can be implemented order by order in perturbation
theory by subtracting the contributions of diagrams with identical
external legs before the sum over $s$ is carried out.
To obtain convergent sums on $s$ the terms coming from positive and
negative $s$ in $S^\pm_{\rm eff}(U)$ have to be lumped together.
There is some ambiguity in this
procedure because one may elect to ``shift'' the $s$-sums
for negative or positive values and be left after the cancelation
with an additional finite part. This ambiguity should reflect
the phase choice freedom discussed before but we have
not traced the correspondence through.

Our purpose here is to replace this perturbative
prescription by a truly non-perturbative formula. The way to proceed
is pretty obvious: Since we have exact homogeneity in the
lattice version on both sides of the defect,
there should be one transfer matrix describing propagation in
the $s$-direction
on one side of the defect and another transfer matrix for the other
side of the defect. Imposing infiniteness in the two directions, projects
out the ground states of the two transfer matrices. The resulting
interface effective action should become
an overlap between the two ground states. Each of these ground states
is the solution to a Hamiltonian problem for fermions
that live in an even dimensional Euclidean space and have nonzero mass.
Once this overlap
formula is obtained, one can forget about the ``wall'', the additional
$s$-dimension, and the bulk infinity. One can build up the argumentation
for the overlap formula itself directly from first principles and
thus circumvent the need to define the infinite $s$-extent by
a limiting procedure.
The purpose of this section is to construct an appropriate
overlap formula for $S_i (U)$.

\subsection 2.1 Transfer matrix from the path integral

The basic problem we wish to solve is to find a way to
replace the Grassmann path integral in equation (2.2)
by matrix products. Since the number of $s$-slices is infinite,
the number of matrices that are multiplied is infinite, and
the full expression has only a formal meaning. However,
the individual factors in the matrix product are perfectly well
defined by the path integral.
Our job is particularly simple because of several choices
we already made in [\ref{Narger}]: We chose to work only
with $1\pm\gamma_{\mu}$ (corresponding to setting the
variable usually denoted by $r$ in the Wilson fermion action
to have absolute magnitude unity); therefore the slices $s$
and $s+1$ will be connected by a transfer matrix and we do not
have to go two steps in $s$. We chose to have the
higher dimensional mass term change only at the ends of one bond
and this will result in a simple overlap formula without necessitating
the insertion of an operator between the two states. Moreover,
our transfer matrix is expected to come out positive definite because
of reflection positivity in the bulk portions; only in this
case will its action as a projector in an infinite product be
an obvious property.

The way to get the transfer matrix
is well known; we choose to follow L\"uscher [\ref{Luesch}].
To start, the Grassmann action has to be written in such
a manner that a natural split of the Grassmann integration
variables between ``holomorphic'' and ``antiholomorphic''
sets is evident. One then proceeds to separate out from the action
the inner product piece that connects different slices; this
piece is kinematical. The left over ``dynamical'' part is then
easily translated into an operator. The operator
is the transfer matrix we are after. There is always some
freedom of grouping together terms into individual factors
in the product of operators; we fix this freedom by
demanding to obtain a formula that only contains an overlap between
states with no operator insertion in the middle.

Since we are only concentrating here on the $s$ dependence of the action
we introduce a compact notation which hides irrelevant dependencies.
We first write the four component Dirac
spinors in terms of two component spinors as follows:
$$\bar\psi_{n,s}=\pmatrix{\bar\chi^L_{n,s}&\bar\chi^R_{n,s}\cr};\ \ \ \ \
\psi_{n,s}=\pmatrix{\chi^R_{n,s}\cr\chi^L_{n,s}\cr}.\eqno\enum$$
The two-component spinors $\chi$ at a fixed $s$ are combined
into a single vector. If $\bar v$ and $u$ are two such vectors, we
define an inner product by
$$(\bar v,u)\equiv \sum_{n\alpha i}\bar v_{n\alpha i}
u_{n\alpha i}.\eqno\enum$$
The sum in the above equation is over the four dimensional space,
$n$; over the two components of the spinor, $\alpha$; and over the
color index, $i$; all at some fixed $s$.
The fermionic action in
(2.1) can be written as
$$\eqalign{S_F(\bar\chi^R,\bar\chi^L,\chi^R,\chi^L,U)=&
-\sum_{s\ge 0} \Bigl[ (\lbhl ,\bp\rphr)+(\rbhr,\bp\lphl)\Bigr]\cr
&-\sum_{s< 0} \Bigl[ (\lbhl,\bm\rphr)+(\rbhr,\bm\lphl)\Bigr]\cr
&+\sum_s \Bigl[ (\lbhl,\bc\lphl)-(\rbhr,\bcd\rphr)\Bigr]\cr
&+\sum_s \Bigl[ (\lbhl,\phr1)+(\bhr1,\lphl)\Bigr].\cr}\eqno\enum$$
$\bpm$ and $\bc$ are operators on the vector space defined above
and they depend on the gauge fields. Explicitly,
$$\bpm_{n\alpha i,m\beta j}=(5\mp
m)\delta_{nm}\delta_{\alpha\beta}\delta_{ij}-{1\over
2}\delta_{\alpha\beta}\sum_\mu \Bigl[ \delta_{m,n+\hat\mu}
U^{ij}_{n,\mu} + \delta_{n,m+\hat\mu}{U^{ji}_{m,\mu}}^*\Bigr]\eqno\enum$$
$$\bc_{n\alpha i,m\beta j}={1\over2}
\sum_\mu \Bigl[ \delta_{m,n+\hat\mu}
U^{ij}_{n,\mu}
- \delta_{n,m+\hat\mu}{U^{ji}_{m,\mu}}^*\Bigr]
\sigma^{\alpha\beta}_\mu\eqno\enum$$
We have used the following representation for the $\gamma$ matrices:
$$\gamma_\mu=\pmatrix{0 & \sigma_\mu\cr \sigma^\dagger_\mu & 0\cr};
\ \ \ \ \ \
\gamma_5=\pmatrix{1 & 0\cr 0  & -1\cr},\eqno\enum$$
where $\sigma_0=i$ and $\sigma_j$; $j=1,2,3$ are the usual Pauli matrices.

The Grassmann path integral in (2.2) can now be converted into
the second quantized operator form using fermion operators
satisfying canonical anti
commutation relations. We follow [\ref{Luesch}] closely and so we will be brief
with some of the intermediate steps relegated to Appendix A. The result for
the effective action, $S_{\rm eff}(U)$, is
$$\eqalign{e^{S_{\rm eff}(U)}=\lim_{s\rightarrow\infty}&
[\det\bm]^{s+{1\over 2}}[\det\bp]^{s+{1\over 2}}\cr
&<b-|\hat D_-\ (\hat T_-)^{s-1}\ (\hat T_+)^{s-1}\
\hat D_+^\dagger|b+>\cr}\eqno\enum$$
$$\hat D_{\pm}=e^{\hat a^\dagger {\rm\bf Q}_{\pm} \hat a}\eqno\enum$$
$$\hat T_{\pm}=e^{\hat a^\dagger {\rm\bf H}_{\pm} \hat a}\eqno\enum$$
$$e^{{\rm\bf Q}_{\pm}}=
\pmatrix{ {1\over\sqrt{\bpm}} & {1\over\sqrt{\bpm}}\bc\cr
0 & {\sqrt{\bpm}}\cr}\eqno\enum$$
$$e^{{\rm\bf H}_{\pm}}=\pmatrix{ {1\over {\bpm}} & {1\over {\bpm}}\bc\cr
\bcd{1\over {\bpm}} & \bcd{1\over {\bpm}}\bc+\bpm\cr}\eqno\enum$$
Since $\bpm$ are both strictly positive (see Appendix A) all the above
equations are well defined. $|b\pm>$ in (2.9) are boundary
conditions at $s=\pm\infty$ respectively. The choice of
these boundary conditions is discussed in the next subsection. $\hat
a_{n A i}$
and $\hat a^\dagger_{n A i}$ are fermion operators satisfying
canonical anti commutation relations:
$$\{\hat a_{nAi},\hat a^\dagger_{mBj}\}
=\delta_{nm}\delta_{AB}\delta_{ij};\ \ \ \ \
\{\hat a_{nAi},\hat a_{mBj}\}=0;\ \ \ \
\{\hat a^\dagger_{nAi},\hat a^\dagger_{mBj}\}=0.\eqno\enum$$
Here the index $A$ and $B$ runs over four  spinor components.
The two transfer matrix operators in (2.10) are hermitian due to
the hermiticity of (2.14). One can compute the determinants of
$e^{{\rm\bf H}_{\pm}}$ and prove that the ${\rm\bf H}_{\pm}$ are
traceless. While this indicates the presence of both
positive and negative eigenvalues the relation between the
two sets of eigenvalues and eigenvectors is not as simple
as it would be in the continuum.

\subsection 2.2 Overlap formula and a phase ambiguity

We still need to supply the boundary conditions appearing in (2.10).
Let us first
absorb the operators $\hat D_-$ and $\hat D^\dagger_+$ into the
boundary states and define new boundary states by
$$|b'\pm>=\hat D^\dagger_\pm|b\pm>.\eqno\enum$$
(2.10) is now given by
$$e^{S_{\rm eff}(U)}=\lim_{s\rightarrow\infty}
[\det\bm]^{s+{1\over 2}}[\det\bp]^{s+{1\over 2}}
<b'-|(\hat T_-)^{s-1}\ (\hat T_+)^{s-1}
|b'+>\eqno\enum$$

It is easy to see that the equations
corresponding to (2.17) for $S^\pm_{\rm eff}(U)$ are
$$e^{S^\pm_{\rm eff}(U)}=\lim_{s\rightarrow\infty}
[\det\bpm]^{2s+1}
<b^\prime \pm|(\hat T_\pm)^{2s-2}
|b^\prime \pm>\eqno\enum$$
$|b^\prime \pm>$ are the modified (similar to (2.16)) boundary states
for the homogeneous cases. The boundary states at $s=\pm\infty$ are the same
for the homogeneous cases. Further, owing to the hermiticity of $\bpm$
and ${\rm\bf H}_\pm$ (c.f.(2.14)) the effective actions for the
homogeneous cases are real. The
result for the effective action at the interface in (2.3) is
$$e^{S_i(U)}=\lim_{s\rightarrow\infty}{
<b'-|(\hat T_-)^{s-1}(\hat T_+)^{s-1}|b'+>\over
\sqrt{<b^\prime -|(\hat T_-)^{2s-2}|b^\prime ->
<b^\prime +|(\hat T_+)^{2s-2}|b^\prime +>}}\eqno\enum$$
Note that the quantities inside the square root are positive.
The limit $s\rightarrow\infty$ projects out the ground states and yields
$$e^{S_i(U)}={<b'-|0-><0-|0+><0+|b'+>\over
|<b^\prime -|0->||<b^\prime +|0+>|}\eqno\enum$$
$|0\pm>$ are the ground states of $\hat T_\pm$ respectively.
In addition to $<0-|0+>$ (2.20) contains two complex factors of modulus
unity; they must be there because, as written, eq. (2.20) seems independent
of the phases of $|0\pm >$. We may just as well admit that there is a
dependence
on the phases and that they are undetermined.
Nothing changes if we make the replacement
$$|b' \pm>~~\rightarrow~~|0\pm>.\eqno\enum$$
We end up with the ``overlap formula'' for the effective action at one
interface,
$$e^{S_i(U)}=<0-|0+>.\eqno\enum$$

The real part of the above equation
is easy to understand.
If we imposed periodic boundary conditions
in the $s$-direction for both the wall and the homogeneous cases
the right hand side of (2.20) will be $|<0-|0+>|^2$. This is
purely real and corresponds to the ``wall''-``anti wall'' case:
One has a finite circular $s$-extent with a step up in the
mass term followed by a step down halfway round the $s$-circle. The
interface action is the sum from the two interfaces and has a
``light'' contribution from pairs of chiral fermions of opposite chirality.

The ambiguity in (2.22) arising from the phases affects only the imaginary part
of $S_i(U)$. Gauge invariance can be broken only by these
phases. Indeed, it was understood for quite a while that the
definition of the real part of the action of chiral
fermions ought to be a trivial matter, it is the imaginary part that
is special to the chiral character reflecting the basic absence of
a parity operator in Minkowski space. Nevertheless, without
exception as far as we know, all older lattice approaches
fail even for the real part. Thus, we feel that some progress
has been made. The real part is defined nonperturbatively and it
is represented by the same object that is supposed to also produce
the imaginary part. It should be stressed that even
if we believe that some chiral gauge theories exist, we still should expect
some difficulty in defining the phase of the chiral determinant in a
gauge invariant approach as long as the construction
itself is insensitive to anomalies.

To make some progress in understanding the
phase problem we extend the overlap formula to the
continuum and compute it in perturbation theory. In perturbation
theory, one should be able to separate the ambiguous and the
unambiguous contributions order by order. In section 3, we
attack the simple problem of perturbation theory in two dimensions and
show that the overlap formula indeed produces the correct anomaly.

\subsection 2.3 First quantized form of the overlap formula

In the last subsection we have reduced the computation of the
effective action to a simple overlap between the ground states of
the transfer matrix operators, $\hat T_+$ and $\hat T_-$. These are
two positive and hermitian operators.
The ground
states of $\hat T_\pm$ are obtained by filling all the states
corresponding to the negative eigenvalues of the
traceless operators ${\rm\bf H}_\pm$
respectively.
Let ${\bf\rm R}$ and ${\bf\rm L}$ diagonalize ${\rm\bf H}_+$
and ${\rm\bf H}_-$ respectively:
$${\bf\rm R}_{\alpha,nAi}{{\bf\rm H}_+}_{nAi,mBj}
{\bf\rm
R}^\dagger_{mBj,\beta}=\lambda^+_{\alpha}\delta_{\alpha\beta}\eqno\enum$$
$${\bf\rm L}_{\alpha,nAi}{{\bf\rm H}_-}_{nAi,mBj}
{\bf\rm
L}^\dagger_{mBj,\beta}=\lambda^-_{\alpha}\delta_{\alpha\beta}\eqno\enum$$
$\lambda^\pm_\alpha$ are the eigenvalues of ${\bf\rm H}_\pm$.
Repeated indices are summed over with the exception of $\alpha$ and $\beta$. We
assume that the diagonalizing matrices are so chosen that the
eigenvalues are arranged in an increasing order. Let us assume that
there are $N$ eigenvalues that are negative for both ${\rm\bf H}_\pm$.
If the numbers are different then the overlap will be zero. The two
ground states are
$$|0+>=
{\bf\rm R}^*_{1,n_1A_1i_1}\cdots
{\bf\rm R}^*_{N,n_NA_Ni_N}
\hat a^\dagger_{n_1A_1i_1}\cdots
\hat a^\dagger_{n_NA_Ni_N}|0>\eqno\enum$$
$$|0->=
{\bf\rm L}^*_{1,n_1A_1i_1}\cdots
{\bf\rm L}^*_{N,n_NA_Ni_N}
\hat a^\dagger_{n_1A_1i_1}\cdots
\hat a^\dagger_{n_NA_Ni_N}|0>\eqno\enum$$
$|0>$ is the vacuum state annihilated by all $\hat a_{nAi}$.
Let $O$ be the restriction of the matrix
${\rm\bf LR}^\dagger$ to the negative eigenvalues of both ${\rm\bf
H}_\pm$.
Then we show in Appendix B that
$$<0-|0+>=\det{O}\eqno\enum$$

\subsection 2.4 The free field case

It is instructive to see the explicit structure of the transfer matrices
in the absence of gauge fields. This would be the first step in setting up
perturbation theory.

Going to Fourier space, for which we maintain a discrete notation
even if the four dimensional volume is infinite, we get the following
forms in a plane wave basis:
$$\left [ e^{{\rm\bf H}_{\pm}}
\right ]^0_{pi,qj}=\delta_{pq}\delta_{ij}
\pmatrix{
{1\over {1\mp m +{1 \over 2} {\hat p}^2}}
& {{i\sigma\cdot{\bar p}} \over {1\mp m +{1 \over 2} {\hat p}^2}}\cr
{{-i\sigma^\dagger \cdot{\bar p}} \over {1\mp m +{1 \over 2} {\hat p}^2}} &
{ {{\bar p}^2} \over { 1\mp m + {1 \over 2} {\hat p}^2 } }  +
1\mp m +{1 \over 2} {\hat p}^2
\cr}\eqno\enum$$
In the above equation only the spinor structure is non-diagonal. As usual
we use $ \bar p_\mu =\sin (p_\mu )$ and
$\hat p_\mu = 2\sin{{p_\mu}\over {2}}$. The eigenvalues
$\lambda$ of the
matrices are the roots of the equation
$$
\lambda + {1\over{\lambda}} = {{1+{\bar p}^2}\over {1\mp m +{1 \over 2} {\hat
p}^2}}+1\mp m +{1 \over 2} {\hat p}^2\eqno\enum$$
and are easily recognized as appearing in the free propagator
obtained in [\ref{Narger}]. The complete free propagator
can be obtained from the explicit form of the free transfer matrices.
Generically, one of the roots is larger than one and the other is its
inverse.

Let us now consider the
eigenvectors corresponding to the above eigenvalues. Following
[\ref{Narger}] we denote $a_\pm = 1\mp m +{1 \over 2} {\hat p}^2 $.
Let the eigenvector(s) corresponding to $\lambda$ be $\psi^\pm_\lambda$
with degeneracy indices suppressed. Write
$$
\psi^\pm_\lambda = \pmatrix{u^\pm\cr v^\pm}\eqno\enum$$
with
$$
i\sigma\cdot{\bar p}v^\pm=(a_\pm \lambda -1) u^\pm~~~~
-i\sigma^\dagger \cdot{\bar p}u^\pm
=({{a_\pm}\over{ \lambda}}-1) v^\pm.\eqno\enum$$
We are interested in the eigenvectors corresponding to $\lambda < 1$.
For $a_-$ we have $a_- / \lambda > 1$ for all momenta and we can use
the second equation, expressing $v$ in terms
of $u$, globally over momentum space. For $a_+$ however a global choice
is impossible. In selecting the definition of the eigenvector
we must check for all locations where the right hand side
has a vanishing prefactor. The identity $(a_+ /\lambda -1) (1-\lambda a_+ )=
{\bar p}^2$ shows that all the trouble spots are at the places
where massless fermions would appear in the na{\" \i}ve
fermion action. Both
factors in the identity cannot vanish simultaneously. It is easy to check
now that the zero momentum point corresponds to the vanishing of
$a_+ / \lambda -1$ while all the other points to the vanishing of
$1-\lambda a_+$. Hence, in a small region around the origin we have to
define the eigenvector by the first equation while outside that
region we should choose the other.  The singling out of the positive side is
reminiscent of  the
fact that only this side carries a Chern-Simons current in the $s$-direction
in the odd dimensional formulation [\ref{Gonsan}].

As we will see in
section 3, the situation in the continuum will be slightly different.
In the continuum equations similar to (2.31) we will have to pick one
equation on the positive side and the other equation on the negative
side for all momenta. We should observe, in view of the above, that
in finite volume formulations on the lattice,
one probably should ascertain that the definition
chosen for the phase is such that it be compatible with a smooth
infinite volume limit.

\subsection 2.5 Chiral determinant and the overlap

We arrived at (2.22) by a rather circuitous route: The first step
is in the paper of Callan and Harvey [\ref{Calvey}]
 where the mass defect is introduced
as a smooth background in a five dimensional theory. Homogeneity
in the fifth direction holds only asymptotically. The second step is
taken in Kaplan's work [\ref{Kaplan}] where the defect is put on the lattice
and
becomes strictly localized in the fifth direction. The third step was taken
in our previous paper [\ref{Narger}] where any five dimensional aspects of the
gauge field were abolished and the lattice action was chosen so as to have
a good transfer matrix in the fifth direction. Also, the defect was chosen
to have the shortest possible extent of just one bond. Clearly,
if (2.22) is right it shouldn't depend on such an indirect derivation and
the simplicity of the expression suggests that it has wider validity, beyond
the
lattice.

Our goal is to provide a simple and direct, albeit formal, argument
for the validity of the overlap formula. Let the chiral Dirac operator
in four Euclidean dimensions be denoted by $X$. The group of
Euclidean space rotations,
$O(4)$, is represented inequivalently in the domain and image of $X$.
Therefore, the usual definition of the determinant needs an extension.
However, both spaces connected by $X$ are Hilbert so the operators
$XX^\dagger$ and $X^\dagger X$ are well defined and have usual determinants.
It is only the phase of $\det (X)$ that really needs to be defined
[\ref{Alvten}].
Since we are only interested in the gauge dependence of this phase
it is sufficient to be able to define $\det (X_0^{-1} X)$ for some fixed
$X_0$ and arbitrary $X$. The latter determinant is of the usual type
and will need regularization; usually this involves some related
eigenvalue problem and this would have been difficult to formulate
for $X$ itself.

Ignoring the infinity of the dimensions
of the spaces we define $\det (X)$ as follows:
Let the space on which $X$
acts be denoted by $V_L$ and the space to which $V_L$ is mapped be
$V_R$. Let $\{ v_L^{(i)} \}$ and $\{ v_R^{(i)} \}$
be orthonormal bases of $V_{L,R}$ respectively. We now define
$\det(X) = \det_{ij} <v_R^{(i)}, X v_L^{(j)}>$. If $Y: V_L \rightarrow V_L$
$\det(XY)=\det(X)\det(Y)$, and replacing $X$ by $X_0$ and $Y$ by $X_0^{-1}X$
we see that ratios will come out correctly.

We now turn to the overlap and show that it is formally related to
the determinant of an operator that has the appearance of an
operator regularized version of $X$. The relationship is formal because
we work
with operators in infinite spaces and we do not worry about the
finiteness of the expressions we are writing down. The final expression
admits a lattice regularization and probably
many others. The lattice regularized version is equation (2.22).

Define a Hermitian traceless operator $H$ in the space $V=V_R \oplus V_L$
$$
H=\pmatrix{m & X^\dagger \cr X & -m }\eqno\enum$$
$H$ is the single particle Hamiltonian of a five dimensional
Minkowski massive Dirac system with real
space identified with Euclidean four space.
Following (2.30) and (2.31) we write:
$$
H\psi_\lambda =\lambda\psi_\lambda;~~~~~~~\psi_\lambda=\pmatrix{u_\lambda
\cr v_\lambda}\eqno\enum$$
$$
X^\dagger v_\lambda =  ( \lambda - m) u_\lambda;~~~~~~~X u_\lambda =
( \lambda +m ) v_\lambda\eqno\enum$$
To get the overlap we do not need the exact eigenvectors: We only
need two sets of linearly independent vectors spanning the $\lambda >0$
subspaces of $V$ for $m>0$ and for $m<0$. This is easily
achieved by observing that the $\lambda$ terms in the above equation
can be replaced by $X$ dependent operators via:
$$
XX^\dagger v_\lambda = (\lambda^2 -m^2 ) v_\lambda ;~~~~~~~
X^\dagger X u_\lambda = (\lambda^2 -m^2 ) u_\lambda\eqno\enum$$
For $m=|m|$ let $u^{(i)}$ be a basis of $V_L$; then the subspace of interest
in $V$ is spanned by
$$
\psi_i^+ =N_i^+ \pmatrix{ u^{(i)} \cr X{1\over {\sqrt {X^\dagger X + m^2}
+|m|}}
u^{(i)}}\eqno\enum$$
Similarly, for $m=-|m|$, let $v^{(i)}$ be a basis of $V_R$; we are now
interested
in the span of
$$
\psi_i^- =N_i^- \pmatrix{  X^\dagger {1\over {\sqrt {XX^\dagger + m^2} +|m|}}
v^{(i)} \cr v^{(i)}}\eqno\enum$$
The overlap is given by
$$
\det_{ij} \left [ 2N_i^- N_j^+ < v^{(i)} , {1\over {\sqrt {XX^\dagger + m^2}
+|m|}} X u^{(j)} >\right ]\eqno\enum$$
This expression can be viewed as a (partially) regularized version
of $\det(X)$ as defined above (for large eigenvalues of $XX^\dagger$ both
the normalization factors tend to ${1\over \sqrt 2}$).
(2.38) will have the same phase as $\det(X)$ for finite square
matrices, if the bases $u^{(i)}$ \& $v^{(i)}$ were chosen to be independent
of $X$; this is quite natural since these bases can be chosen before $X$ is
given.
Also, note that the choice of a different pair of equations from (2.34)
and (2.35) was instrumental and was justified by demanding that the
expressions (2.36) and (2.37) be valid even when $X$ or $X^\dagger$ have
zero modes. This is reminiscent of (2.31) and the discussion thereafter.

The regularization of the second quantized system built from $H$ can be carried
out gauge invariantly using the gauge invariant single particle energies.
On the lattice one loses the exact match between the two cases $m=\pm |m|$.
As a matter of fact, for an arbitrary background
gauge field there is no guarantee that there will be equal numbers
of single particle states in both states making up the overlap and
therefore the latter can easily vanish. It is in this way that
we expect the lattice to correctly reproduce instanton effects. Note
that the two transfer matrices provide, via the dimensionalities
of the appropriate subspaces a new definition of the topological
charge associated with
a lattice configuration of gauge fields.
This definition is manifestly gauge invariant
and naturally integer valued.
More precisely, the topological charge should be identified
with the signed number of level crossings through eigenvalue unity
when the mass term is changed from positive to negative. Explicitly,
$$
n_{\rm top} = {1\over 2} Tr \left ( {{\bf {H_+}}\over {\sqrt{{\bf {H_+}}^2}}} -
{{\bf {H_-}}\over {\sqrt{{\bf {H_-}}^2}}} \right )\eqno\enum$$
As expected with any lattice definition, there are ``singular'' gauge field
configurations for which the topological charge is not defined; this
permits the integer to change when the background gauge fields are smoothly
deformed. One can smoothly deform gauge fields from one topological
class to another because the lattice has wiped out the manifold structure of
spacetime
and the space of gauge transformations is no longer disconnected. In this
case the ``singular gauge fields'' are the ones for which ${\bf {H_\pm}}$ have
zero
eigenvalues. This can't happen for gauge fields sufficiently close
to the identity in view of our study of the free case. Note that in a
perturbative
expansion around the free case the finite radius of convergence of the series
will
be determined by the first encountered ``singular'' gauge configuration.

It should be no surprise that we unintentionally obtained a lattice definition
of
$n_{\rm top}$: the dynamics of chiral fermions has a universal sensitivity to
the
topological charge of the gauge background and any respectable approach to
the regularization of chiral gauge theories must have some definition of
$n_{\rm top}$ hidden in it [\ref{Topol}].

\section 3. Two dimensional anomaly from the continuum overlap formula

The formal continuum limits of $\bH$ in (2.14)
correspond to Hamiltonians of massive Dirac fields
coupled to a gauge field with the masses being $\pm m$ respectively
\footnote{${}^\dagger$}{In this section $m\equiv |m|$.}.
To obtain this, one uses the standard representation for $U_{n,\mu}$
and expands the right-hand side of (2.14) in powers of the lattice
spacing, $a$. The leading order term gives the continuum Hamiltonians.
The lattice overlap is replaced by the overlap between the
ground states of these two  Dirac Hamiltonians. This introduces a
significant simplification in the algebra.
In this section, starting
from the Hamiltonians in two dimensions
and using Schr\"odinger perturbation theory,
we show that the overlap produces the correct consistent anomaly
for the case of Abelian gauge field in the limit where
$m\rightarrow\infty$. In this limit
the other modes become infinitely heavy and
we expect to be left with only the interesting piece,  namely the
zero mode attached to the abrupt mass defect.

\subsection 3.1 Problem definition

The continuum limits of $\bH$, in momentum space, are
$$-{\bH}(p,q)=H^\pm_0(p)\delta_{p,q}+\sum_k z_k\tau\delta_{q,p+k}
+\sum_k z^*_k\tau^\dagger\delta_{p,q+k}\eqno\enum$$
$$H^\pm_0(p)=\pmatrix {\mp m & p_1-ip_2\cr p_1+ip_2 & \pm
m\cr}\eqno\enum$$
$$\tau=\pmatrix{0 & 0\cr 1 & 0\cr} \eqno\enum$$
$$z_k=\int d^2x e^{ikx} [A_1(x) + i A_2(x)]\eqno\enum$$
The above Hamiltonians are the Hamiltonians associated with the
continuum limit of the action, (2.1), with the coefficient of the
last term being $(5\mp m)$. The $s$-direction plays the role of
Euclidean time.
The two
Dirac matrices are the Pauli matrices $\sigma_1$ and $\sigma_2$.
$(1\pm\sigma_3)$ projects onto right and left chiral states.
In the above equations, $p,q$ and $k$ are two-component momenta.
$\sum_k$ is a notation for the momentum integral ${1\over
(2\pi)^2}\int d^2k$. $A_1(x)$ and $A_2(x)$ are the two components of
the Abelian gauge field.

The unperturbed Hamiltonians are diagonal in momentum space. For each
momentum, the unperturbed Hamiltonians have two eigenstates one with
positive
energy and one with negative energy. We denote these two unperturbed
states by $\psi^\pm_p$ (positive energy) and $\chi^\pm_p$ (negative
energy). The states are labeled by their momentum.
Let
$$-\bH\Psi^\pm_p=\Lambda^\pm_p\Psi^\pm_p;\ \ \ \
\Lambda^\pm_p > 0\eqno\enum$$
be the set of eigenvectors of the full Hamiltonians corresponding to positive
eigenvalues. The unperturbed limit of $\Psi^\pm_p$ is $\psi^\pm_p$,
hence the labeling of the states by $p$.
The overlap matrix is defined as
$$O_{pq}={\Psi^-_p}^\dagger \Psi^+_q\eqno\enum$$
and the effective action at the interface is given by
$$e^{S_i(A)}=\det O\eqno\enum$$
This is the continuum limit of (2.22).
We now proceed to compute $S_i(A)$ in perturbation theory. We assume
that there is no zero momentum term in the gauge field; i.e,
$z_0=0$. If $z_0\ne 0$, it can be combined with the unperturbed
Hamiltonian and will amount to a shift in the momenta.

\subsection 3.2 Eigenvectors and eigenvalues of $H^\pm_0$

$H^\pm_0$ have the same set of eigenvalues with the positive and
negative eigenvalues occurring in pairs.
$$\eqalign{
H^\pm_0\psi^\pm_p&=\lambda_p\psi^\pm_p\cr
H^\pm_0\chi^\pm_p&=-\lambda_p\chi^\pm_p\cr};\ \ \
\lambda_p > 0\eqno\enum$$
The eigenvectors are labeled by a momentum index and are
given by
$$\psi^+_p(q)={1\over N(p)}\pmatrix {p_1-ip_2\cr
\lambda_p+m\cr}\delta_{pq};\ \ \ \
\psi^-_p(q)={1\over N(p)}\pmatrix {\lambda_p+m\cr
p_1+ip_2\cr}\delta_{pq}\eqno\enum$$
$$\chi^+_p(q)={1\over N(p)}\pmatrix {-(\lambda_p+m)\cr
p_1+ip_2\cr}\delta_{pq};\ \ \ \
\chi^-_p(q)={1\over N(p)}\pmatrix {-(p_1-ip_2)\cr
\lambda_p+m\cr}\delta_{pq};\ \ \ \
\eqno\enum$$
$$\lambda_p=\sqrt{|p|^2+m^2} > 0;\ \ \ \
N(p)=\sqrt{2\lambda_p(\lambda_p+m)}\eqno\enum$$
The eigenvectors are globally smooth in the momentum index $p$.
Both sets, $\{\psi^+_p,\chi^+_p\}$
and $\{\psi^-_p,\chi^-_p\}$, are orthonormal and complete.
$$\eqalign{
\sum_k [\psi^\pm_p(k)]^\dagger \psi^\pm_q(k) &= \delta_{pq}\cr
\sum_k [\chi^\pm_p(k)]^\dagger \chi^\pm_q(k) &= \delta_{pq}\cr
\sum_k [\psi^\pm_p(k)]^\dagger \chi^\pm_q(k) &= 0\cr}\eqno\enum$$
The set $\{\psi^-_p,\chi^-_p\}$ is related to $\{\psi^+_p,\chi^+_p\}$
by
$$\eqalign{
\psi^-_p=\Sigma_3\chi^+_p; \ \ \ & \ \ \ \chi^-_p=\Sigma_3\psi^+_p\cr
\Sigma_3(q,k)& = \sigma_3\delta_{qk}\cr}\eqno\enum$$
where $\sigma_3$ is the third Pauli matrix. The overlap matrices
between the eigenstates of $H^+_0$ and $H^-_0$ are
$$[\psi^-_p]^\dagger \psi^+_q = [\chi^+_p]^\dagger\chi^-_q=
{p_1-ip_2 \over \lambda_p}\delta_{pq}\equiv P_{pq}\eqno\enum$$
$$[\chi^-_p]^\dagger \psi^+_q =- [\psi^-_p]^\dagger\chi^+_q=
{m\over \lambda_p}\delta_{pq}\equiv Q_{pq}\eqno\enum$$
The above equations follow from (3.9) and (3.10).

\subsection 3.3 The overlap formula in perturbation theory

In perturbation theory,
the anomaly in two dimensions should show up at second order. To
second order the eigenvectors corresponding to the positive
eigenvalues will have the following form:
$$\eqalign{
\Psi^+_p & = \psi^+_p + \psi^+_q A_{qp} + \chi^+_q B_{qp}
+ \psi^+_q W_{qp} + \chi^+_q X_{qp}\cr
\Psi^-_p & = \psi^-_p + \psi^-_q C_{qp} + \chi^-_q D_{qp}
+ \psi^-_q Y_{qp} + \chi^-_q Z_{qp}\cr}\eqno\enum$$
$A,B,C$ and $D$ are coeffecients linear in the gauge field and
$W,X,Y$ and $Z$ are coeffecients quadratic in the gauge field.
Summation over repeated indices is implied.
The overlap matrix, (3.6), to second order in the gauge field is
$$\eqalign{
 O= & P + PA - QB +C^\dagger P + D^\dagger Q + PW -QX + Y^\dagger P \cr
 & + Z^\dagger Q + C^\dagger PA - C^\dagger QB + D^\dagger QA + D^\dagger
P^\dagger B\cr} \eqno\enum$$
In deriving the above equation from (3.16), we have used (3.12),
(3.14) and (3.15).
The effective action, $S_i(A)$, to second order in the gauge field is
$$\eqalign{S_i(A)- S_i(0) & = \ln\det P^{-1}O\cr
& = \Tr\ln P^{-1}O\cr
& = T_{11}+T_{12}+T_{21}+T_{22}+T_{23}+T_{24}\cr}\eqno\enum$$
$$\eqalign{
T_{11} = & \Tr (A+C^\dagger)\cr
T_{12} = & \Tr P^{-1}Q (D^\dagger - B)\cr
T_{21} = & \Tr (W+Y^\dagger)\cr
T_{22} = & -{1\over 2} \Tr (A^2 +{C^\dagger}^2)\cr
T_{23} = & -{1\over 2} \Tr (P^{-1}QB)^2 -{1\over 2}\Tr
(P^{-1}QD^\dagger)^2 + \Tr P^{-1}D^\dagger P^{-1}Q^2B \cr
& + \Tr P^{-1}D^\dagger P^\dagger B +\Tr P^{-1}QBA - \Tr P^{-1}Q
C^\dagger D^\dagger\cr
T_{24} = & \Tr P^{-1}Q (Z^\dagger - X)\cr}\eqno\enum$$
$T_{11}$ and $T_{22}$ are linear in the gauge field and the rest are
quadratic in the gauge field. We have used the fact that $P$ and $Q$
are diagonal (c.f.(3.14) and (3.15)).

\subsection 3.4 First and second order perturbation theory

The perturbing term is the same for both the Hamiltonians, $\bH$,
(c.f.(3.1)):
$$H_1(p,q)=\sum_k z_k\tau\delta_{q,p+k} + \sum_k
z^*_k\tau^\dagger\delta_{p,q+k}.\eqno\enum$$
The eigenvalues, $\Lambda^\pm_p$, in (3.5) have a perturbation
expansion of the form
$$\Lambda^\pm_p=\lambda_p+{\lambda^\pm_p}^
{(1)}+{\lambda^\pm_p}^{(2)}\eqno\enum$$
${\lambda^\pm_p}^{(1)}$ is linear in gauge field and
${\lambda^\pm_p}^{(2)}$
is quadratic in gauge field. Standard Schr\"odinger perturbation
theory gives
$${\lambda^\pm_p}^{(1)}={\psi^\pm_p}^\dagger H_1 \psi^\pm_p\eqno\enum$$
$$A_{pq}={{\psi^+_p}^\dagger H_1 \psi^+_q\over
\lambda_q-\lambda_p};\ \ \ \ p\ne q\eqno\enum$$
$$B_{pq}={{\chi^+_p}^\dagger H_1 \psi^+_q\over
\lambda_q+\lambda_p}\eqno\enum$$
$$C_{pq}={{\psi^-_p}^\dagger H_1 \psi^-_q\over
\lambda_q-\lambda_p};\ \ \ \ p\ne q\eqno\enum$$
$$D_{pq}={{\chi^-_p}^\dagger H_1 \psi^-_q\over
\lambda_q+\lambda_p}\eqno\enum$$
The diagonal terms, $A_{pp}$ and $C_{pp}$, are not determined above.
Imposing orthonormality of the eigenvectors,
$${\Psi^\pm_p}^\dagger \Psi^\pm_q =\delta_{pq},\eqno\enum$$
to first order, yields
$$A_{pp}+A^*_{pp}=0;\ \ \ \ C_{pp}+C^*_{pp}=0;\ \ \ \forall
p.\eqno\enum$$
The imaginary parts of $A_{pp}$ and $C_{pp}$ remain undetermined and
this arbitrariness is the phase ambiguity in the eigenvectors and the
overlap. One choice to fix this ambiguity is to impose the condition
that the overlap between the true eigenstate and the unperturbed one, namely
${\Psi^\pm_p}^\dagger\psi^\pm_p$, be real. This condition corresponds
to Brillouin--Wigner perturbation theory.

Since $H_1$ is hermitian (c.f.(3.20)) it follows from (3.23) and (3.25) that
$$A^\dagger = -A;\ \ \ \ C^\dagger=-C.\eqno\enum$$
This is true for any choice of fixing the ambiguity. $\Sigma_3$
defined in (3.13) has the property
$$\Sigma_3^\dagger H_1 \Sigma_3 = -H_1\eqno\enum$$
This along with the hermitian nature of $H_1$ relates $B$ and $D$ in
(3.24) and (3.26) by
$$D=-B^\dagger.\eqno\enum$$

Second order perturbation theory is needed to evaluate $T_{21}$ and
$T_{24}$.
Owing to the diagonal nature of $P$ and $Q$ (c.f.(3.14) and (3.15)),
we only need the diagonal terms of $W,X,Y$ and $Z$. The diagonal terms
of $W$ and $Y$ are not determined by the eigenvector condition. Enforcing
the orthonormality, (3.27), to second order, yields
$$\eqalign{
W_{pp}+W^*_{pp}& =-\sum_q\Bigl[ A^\dagger_{pq}A_{qp}+
B^\dagger_{pq}B_{qp}\Bigr]\cr
Y_{pp}+Y^*_{pp}& =-\sum_q\Bigl[ C^\dagger_{pq}C_{qp}+
D^\dagger_{pq}D_{qp}\Bigr]\cr}\eqno\enum$$
As with $A$ and $C$, the imaginary parts of $W_{pp}$ and $Y_{pp}$ remain
undetermined. Brillouin--Wigner perturbation theory sets all of them
to zero:
$$A_{pp}=C_{pp}=W_{pp}=Y_{pp}=0.\eqno\enum$$

For $X$ and $Z$, second order perturbation theory gives
$$\eqalign{
2\lambda_p X_{pp}=\sum_{q\ne p}\Bigl[
(\lambda_p+\lambda_q) &
B_{pq}A_{qp}+(\lambda_p-\lambda_q)C_{pq}B_{qp}\Bigr]\cr
&+\Bigl[2\lambda_pA_{pp}-{\lambda^-_p}^{(1)}-{\lambda^+_p}^
{(1)}\Bigr]B_{pp}\cr}
\eqno\enum$$
$$\eqalign{
2\lambda_p Z_{pp}=\sum_{q\ne p}\Bigl[
(\lambda_p+\lambda_q) &
D_{pq}C_{qp}+(\lambda_p-\lambda_q)A_{pq}D_{qp}\Bigr]\cr
&+\Bigl[2\lambda_pC_{pp}-{\lambda^-_p}^{(1)}-{\lambda^+_p}^
{(1)}\Bigr]D_{pp}\cr}
\eqno\enum$$

\subsection 3.5 Imaginary part of the effective action

Using the results of the previous subsection, we now compute the
imaginary part of the effective action given by (3.18). We define
$$I_1={1\over 2}(T_{11}-T^*_{11});\ \ \ \ I_2={1\over 2}
(T_{21}-T^*_{21}).\eqno\enum$$
Because they only involve the imaginary parts of
$A_{pp},C_{pp},W_{pp}$ and $Y_{pp}$, they are both completely
arbitrary.

{}From (3.9), (3.10), (3.20), (3.24) and (3.26) it is clear that
$$B_{pp}=D_{pp}=0\eqno\enum$$
if $z_0=0$. Therefore
$$T_{12}=0$$
in (3.19). Because of (3.29), $T_{22}$ in (3.19) is real and does not
contribute to the imaginary part of $S_i(A)$.

{}From (3.14) and (3.15) it follows that
$$P^\dagger = P^{-1}(1-Q^2).\eqno\enum$$
This, along with (3.29) and (3.31), reduces $T_{23}$ in (3.19) to
$$T_{23}=-\Tr (P^{-1}QB)^2 - \Tr (P^{-1}B)^2 + \Tr
P^{-1}Q(BA-CB)\eqno\enum$$
Using (3.29), (3.31) and (3.37), (3.34) and (3.35) yield
$$Z^*_{pp}-X_{pp}=\sum_q {\lambda_q\over \lambda_p}
(C_{pq}B_{qp}-B_{pq}A_{qp})\eqno\enum$$
Using (3.14), (3.15) and (3.40), $T_{24}$ in (3.19) is given by
$$T_{24}=\sum_{pq} {m\over p_1-ip_2}{\lambda_q\over \lambda_p}
(C_{pq}B_{qp}-B_{pq}A_{qp})\eqno\enum$$

To proceed further with $T_{23}$ and $T_{24}$ we need the explicit
expressions for $A,B$ and $C$. The defining equations for $A,B$ and
$C$ are (3.23), (3.24) and (3.25). Using the eigenvectors given by
(3.9) and (3.10), and $H_1$ given by (3.3), (3.4) and (3.20), we get
$$A_{pq}=z_{q-p}a(p,q)-z^*_{p-q}a^*(q,p)\eqno\enum$$
$$B_{pq}=z_{q-p}b(p,q)-z^*_{p-q}d(p,q)\eqno\enum$$
$$C_{pq}=-z_{q-p}a(q,p)+z^*_{p-q}a^*(p,q)\eqno\enum$$
$$a(p,q)={(\lambda_p+m)(q_1-iq_2)\over
(\lambda_q-\lambda_p)N(p)N(q)}\eqno\enum$$
$$b(p,q)={(p_1-ip_2)(q_1-iq_2)\over
(\lambda_q+\lambda_p)N(p)N(q)}\eqno\enum$$
$$d(p,q)={(\lambda_p+m)(\lambda_q+m)\over
(\lambda_q+\lambda_p)N(p)N(q)}\eqno\enum$$
Using the above equations for $A,B$ and $C$ along with (3.14) and
(3.15), the imaginary part of $T_{23}$ in (3.39) and the imaginary part
of $T_{24}$ in (3.41) give
$$\eqalign{I_3 &= {1\over 2} (T_{23}+T_{24}-T^*_{23}-T^*_{24})\cr
& = \sum_{p} \Bigl[
k(p)z_{p}z_{-p}-k^*(p)z^*_{p}z^*_{-p}\Bigr]\cr}\eqno\enum$$
$$k(p)=\sum_q c(p+q,q)
{1\over
[p_1+q_1+i(p_2+q_2)][q_1+iq_2]}\eqno\enum$$
$$c(p,q)=
{m(\lambda_p\lambda_q+\lambda^2_p+\lambda^2_q-m^2)\over
4\lambda_p\lambda_q(\lambda_p+\lambda_q)}
\eqno\enum$$
The above two equations have a semblance to the ``bubble'' diagram in
ordinary Feynman diagram language. $p$ is the external momenta. The
second factor in (3.49) is the product of the two chiral fermion
propagators occurring in the bubble. $c(p,q)$ is a
regulator for high momenta introduced by the overlap definition of the
effective action.

Finally,
$${\rm Im}[S_i(A)-S_i(0)]= I_1+I_2+I_3\eqno\enum$$
The first two terms are ambiguous and the last term is completely
determined. It is worthwhile noting here that the condition $z_0=0$ is
not needed to remove all arbitrariness in $I_3$. This is because the
arbitrariness in the last terms in (3.34) and (3.35) is exactly
cancelled by the arbitrariness in the last term in (3.39).

\subsection 3.6 Anomaly

We now want to check if (3.51) produces the correct anomaly. The
anomaly arises due to non-local contributions to $I_3$. The
arbitrariness in $I_1$ and $I_2$ arising from choices for the diagonal
terms of $A,C,W$ and $Y$ should be local and cannot affect the
anomaly.

We start by showing that $I_3$ in (3.48) has a non-local piece. Note
that $k(p)$ defined in (3.49) is finite. It can be rewritten as
$$k(p)={1\over p_1+ip_2}\sum_q {1\over q_1+iq_2}
[c(q+p,q)-c(q-p,q)].\eqno\enum$$
We now expand $[c(q+p,q)-c(q-p,q)]$ in powers of $p$. From (3.50), we
find
$$c(q+p,q)-c(q-p,q)= -{3m|q|^2\over
8\lambda_q^5}(q_1p_1+q_2p_2)+O(p^2)\eqno\enum$$
Using (3.53) in (3.52) gives
$$k(p)=-{p_1-ip_2\over p_1+ip_2}\sum_q {3m|q|^2\over
16\lambda_q^5}+O(|p|/m)\eqno\enum$$
The integral in the above equation is finite. When
$m\rightarrow\infty$, we get the following non-local expression for
$I_3$ in (3.48):
$$I_3={1\over 16\pi} \sum_p\Bigl[
{p_1+ip_2\over p_1-ip_2}z^*_pz^*_{-p}
-{p_1-ip_2\over p_1+ip_2}z_pz_{-p}\Bigr]\eqno\enum$$
Note that only in the limit $m\rightarrow\infty$ all the additional
fields become infinitely massive and we are left with one chiral fermion.

Now we focus on $I_3$ to extract the anomaly [\ref{Jackiw}]. The
anomaly is defined as
$${\cal A}(x)=\sum_\mu \partial_\mu {\delta I_3\over \delta
A_\mu(x)}\eqno\enum$$
The momentum space equivalent of the above equation using (3.4) is
$$\eqalign{
i {\cal A}_p&=\Bigl[ p{\delta I_3\over \delta z_{-p}}+p^*
{\delta I_3\over \delta z^*_p}\Bigr]\cr
& = {1\over 8\pi}\Bigl[
(p_1+ip_2)z^*_{-p}-(p_1-ip_2)z_{p}\Bigr]\cr}\eqno\enum$$
We have used (3.55) in deriving the last line above.
{}From (3.4) we have
$$z_p=A_{1,p}+iA_{2,p};\ \ \ \ z^*_{-p}= A_{1,p}-iA_{2,p}.\eqno\enum$$
Using (3.58), the anomaly, (3.57), becomes
$${\cal A}_p= {1\over 4\pi}(p_2A_{1,p}-p_1A_{2,p}) \eqno\enum$$
Since the field strength in two dimensions is
$$F_{12}(p)=i(p_2A_{1,p}-p_1A_{2,p})\eqno\enum$$
the anomaly, (3.59), can be rewritten as
$${\cal A}_p=-{i\over 4\pi}F_{12}(p)\eqno\enum$$
The normalization of the right hand side tells us that, although we
computed only the abelian anomaly, we still can conclude that it is in
the consistent rather than the covariant form \footnote{${}^\dagger$}{
In the covariant case the factor of ${1\over 4\pi}$ would be replaced
by ${1\over 2\pi}$}.

\subsection 3.7 Consistent versus covariant anomalies

Once we have an explicit formula for the effective action induced by
chiral fermions the currents defined by varying this action with
respect to the external gauge fields will, by construction, have
anomalies in the ``consistent'' form at the expense of covariance under
gauge transformations [\ref{Covcon}].
 However, it is known that in the case that
the defect is string--like rather than wall--like (i.e. the codimension
of the defect is two rather than one) that the anomaly one obtains from
the Callan-Harvey arguments [\ref{Calvey}] has the covariant form. This was
emphasized by Naculich [\ref{Covcon}].
It also is true of the wall case. Thus, in our case, the Callan-Harvey
analysis cannot fully account for the anomaly, in particular the result
obtained in eq. (3.61) above. This is true even allowing for
the additional changes induced
by the lattice regularization of the five dimensional calculation
of the induced Chern-Simons term, namely its appearance on only one side
of the defect [\ref{Gonsan}].

For simplicity, we shall again restrict ourselves
to two dimensions. To see
the difference between consistent and covariant forms and not have to rely
on normalizations we have chosen to work here with the non-abelian case.
Let us start by explaining how
Callan and Harvey end up with the covariant form of the anomaly. We use a
compact differential form notation that has proven useful in
studying anomalies in arbitrary dimensions and the relations
between them [\ref{Zumino}].

Let ${\cal M}$ be a three dimensional manifold with boundary.
$A$ is a three dimensional vector potential written as a Lie Algebra
valued one form. $F$ is the
associated curvature $F=dA+A^2$. $tr$ denotes trace
over representation indices.
The three dimensional Chern-Simons three--form is given by
$$
\Omega = tr [ AdA +{2\over 3} A^3 ]\eqno\enum$$

We choose the Chern-Simons action as $S_{CS} =-{{i}\over{4\pi}}
\int_{\cal M} \Omega$ with
no boundary terms added.
Under an arbitrary variation $\delta$ we have
$$
\delta S_{CS} = -{{i}\over{2\pi}}\int_{\cal M} tr (\delta A F )
-{{i}\over{4\pi}}\int_{\partial {\cal M}}
tr (\delta A A )\eqno\enum$$
Restricting the variation to a gauge variation $\delta A={\delta}_h A
\equiv dh +[A,h]$ the first integrand becomes the $d$ of something and
we obtain
$$
{\delta}_h S_{CS} = -{{i}\over{4\pi}}\int_{\partial {\cal M}} tr
(hdA)\eqno\enum$$
The boundary of the manifold is two dimensional and restricting
the gauge fields to it we obtain the two dimensional anomaly.
Clearly $dA$ does not transform covariantly because the term $A^2$
is missing.

Suppose we have the Callan-Harvey wall set-up and $x_3 \equiv s$ is
the third direction. We view the system as three dimensional and
define the three dimensional currents (more precisely, their Hodge duals)
${\cal J} = {{\delta S_{CS }}\over
{\delta A}}$. Far away from the defect, at
$x_3 \rightarrow \pm \infty$,
the integration of the massive fermions is presumed to induce
$\pm {1\over 2}S_{CS}$
terms. \footnote{${}^\dagger$}{The reader
should not worry about overall normalizations,
because
we only want to trace the difference between the two forms of the anomaly.}
Note however that the computations leading to this assumption are
most naturally defined for a homogeneous system on a boundary
free manifold in the limit it becomes infinite [\ref{Redlich}]. While we cannot
write explicit formulae for ${\cal J}$ everywhere, we do know
the currents in the asymptotic regime. Specialize now to a background
which is two dimensional and $s$
independent. We then have, as $s\rightarrow \infty$,
${\cal J}_3=-{{i}\over{4\pi}}F_{12}$
and, as $s\rightarrow -\infty$,
${\cal J}_3={{i}\over{4\pi}}F_{12}$;
the $1$ and $2$
components of ${\cal J}$ vanish in the asymptotic regions because of
our choice of background.

Because of gauge invariance and absence of three dimensional anomalies
$$
D_\mu {\cal J}_\mu =0\eqno\enum$$
everywhere. $D_\mu$ is the covariant
derivative in the $\mu^{\rm th}$
direction. Integrating the equation over s from $-\infty$ to $\infty$
we get
$$
D_i J_i = {{i}\over{2\pi}}F_{12}\eqno\enum$$
In (3.65) $\mu =1,2,3$ and in (3.66) $i=1,2$ and $J_i = \int_s {\cal J}_i
(x_1 x_2 x_3 =s )$.
We have obviously obtained the covariant form of the anomaly
\footnote{${}^\dagger$}{The nonabelian field strength appears rather than
$\epsilon_{ij}\partial_jA_i$. The prefactor is ${1\over 2\pi}$ rather
than ${1\over 4\pi}$.}.
This means that
the currents $J_i$ can't be written as the variation of something; this
is a bit strange because the ${\cal J}$ were defined as a variation. The
problem
was introduced when we evaluated ${\cal J}$: we tacitly assumed that
the variation $\delta A$ vanishes at infinity (the boundary of our manifold)
and therefore ignored the boundary term in (3.63). This procedure
is correct for the three dimensional current in the bulk but wrong at
infinity.
However, the justification for the Chern-Simons form of the induced action
holds
only in bulk. To get the consistent form of the anomaly we have seen above
that one needs to know that the Chern-Simons form of the induced actions
holds also on the boundary of the manifold.
We do not know this to be true but we could easily accept that a carefully
chosen boundary condition at $s=\pm\infty$ will make this happen.
We must assume that this choice of boundary conditions has been carried
out if we want $S_i (U)$ to behave properly.

The additional piece of current that would
turn the above defined $J_i$ into a variation of an action resides
at the boundaries at $s=\pm\infty$. It is easy to see that it is
given by $\Delta J_i = {{i}\over{4\pi}}\epsilon_{ij} A_j$ and that the new
current
$J_i +\Delta J_i$ has the consistent anomaly.
In summary, to get the right form for the anomaly one must
ensure that the boundary conditions at infinity are as undisruptive as
possible.  This is reminiscent
of the boundary conditions needed in defining the APS index theorem
for manifolds with boundary [\ref{Index}].

We now proceed to propose an overlap formula for the
covariant currents. The formula has some
geometric appeal. In computing the anomaly of these currents
we shall realize that the difference between them and the previously
$s$-integrated ${\cal J}_i$'s is due to higher energy excitations in the
massive
fermion systems. Na{\"\i}vely one would have ignored such excitations
because of suppression by large energy denominators, but, they are
sufficiently numerous to make a difference. This formula
might be particular useful when defining currents associated
with global anomalous symmetries\footnote{${}^\dagger$}
{Recent studies of a global anomaly in a theory regularized by
infinitely many Pauli-Villars fields have given encouraging results
[\ref{Aoki}]. } in an anomaly--free gauge theory [\ref{Banks}].

Our main purpose is to identify  the terms in the variation of
the overlap that produce
the $\Delta J_i$ contribution.
Let $A_\mu^{\rm ext} (x)$ be an external
vector potential coupled to the
two Hamiltonians that would correspond to a symmetry whose current
we wish to calculate.
Make the external vector potential infinitesimal; a current
with the consistent form of the anomaly would emerge if we
evaluated ${{\delta (<0-|0+>)}\over {<0-|0+>}}$. The variation of
the states $|0\pm >$ contains a piece proportional to the state itself
and additional contributions
from excited states. If we neglect the excited states we obtain for the
candidate covariant current an expression
made out of two contributions, one from
each side of the overlap: $<0+|\delta 0+> + <\delta 0-|0->$. More
explicitly, we have:
$$
J_\mu^{\rm cov} (x) = <0+|{{\partial}\over {\partial A_\mu^{\rm ext} (x)}}
|0+>-<0-|{{\partial}\over {\partial A_\mu^{\rm ext} (x)}}
|0->\eqno\enum$$

Each one of the terms in the above equation has a geometrical
interpretation: Remember that the phases of $|0\pm>$ are ambiguous to some
extent. The related $U(1)$ gauge symmetry over the space
of vector potentials has a naturally associated
connection [\ref{Uhol}], and the
current is written as the difference between these
two connections. Under a $U(1)$ gauge transformation the phases of the
states change but this should leave the anomaly unchanged as long
as the  gauge change is local. So the anomaly is a $U(1)$ gauge invariant
(under a class of gauge functions restricted by
locality) and should reside in the curvature of the connection. Of course,
had we insisted that the current be a total derivative there would
have been no curvature. It so happens
that all the covariant anomaly is contained in the curvature. We really need
only the difference between the two curvatures, denoted by ${\cal F}$:
$$
{\cal F}_{\mu x , \nu y} = {{\partial J_\mu^{\rm cov} (x)}
\over {\partial A_\nu^{\rm ext} (y)}}-
{{\partial J_\nu^{\rm cov} (y)}
\over {\partial A_\mu^{\rm ext} (x)}}\eqno\enum$$
When ${\cal F}$ is integrated over a two dimensional disk embedded
in the space of vector potentials it gives the difference
between the Berry phases [\ref{Berry}] accumulated
when each one of the vacua is adiabatically moved around the circumference
of the disk. The unremovability of these phases is related to the
presence of degeneracy submanifolds in the space of vector potentials.

Let us sketch the perturbative evaluation of ${\cal F}$ in an
abelian two dimensional situation. The calculation has been first carried
out by Niemi et. al. [\ref{Nienwu}]. Since we need the currents only to
first order in the gauge fields ${\cal F}$ is needed only to zeroth order.
Using intermediate states and going to first quantized formalism
we have
$$
{\cal F}_{\mu x , \nu y}= {\cal F}_{\mu x , \nu y}^+-{\cal F}_{\mu x , \nu y}^-
\eqno\enum$$
$$
\eqalign{
{\cal F}_{\mu x , \nu y}^\pm =&-
\sum_{p,q} {1\over (\lambda_p+\lambda_q)^2}
\big \{ \big [
<\psi_p^\pm |{{\partial \bH}\over {\partial A_\mu^{\rm ext} (x)}}|\chi_q^\pm >
<\chi_q^\pm |{{\partial \bH}\over {\partial A_\nu^{\rm ext} (y)}}|\psi_p^\pm
>\cr
-&
<\psi_p^\pm |{{\partial \bH}\over {\partial A_\nu^{\rm ext} (y)}}|\chi_q^\pm >
<\chi_q^\pm |{{\partial \bH}\over {\partial A_\mu^{\rm ext} (x)}}|\psi_p^\pm >
\big ]\cr
 -& \big [ \mu x ~~\leftrightarrow ~~\nu y \big ] \big \} }
\eqno\enum$$

Here we have used the notations from section (3.1). Plugging in explicit
formulae
one gets after some algebra:
$$
{\cal F}_{\mu x , \nu y}=4m\epsilon_{\mu\nu}\sum_{k,k'}
{{e^{i(k^\prime - k)\cdot (x-y)}}\over{\lambda_k\lambda_{k^\prime} (\lambda_k
+\lambda_{k^\prime} )}}\eqno\enum$$
The
structure is reminiscent of equation (3.50).
The following identity converts the momentum
integral to a three dimensional integral associated with a bubble
diagram:
$$
\eqalign{
\int &{{d^2 p}\over { (2\pi )^2}}{{1}\over{\lambda_{p+q/2} \lambda_{p-q/2}
 (\lambda_{p+q/2}+\lambda_{p-q/2} )}} =\cr 2
\int & {{d^3 p}\over { (2\pi )^3}}{{1}\over{(p_0^2 + (p+q/2)^2 +m^2 )(
p_0^2 +(p-q/2 )^2 +m^2 )}}
}\eqno\enum$$
In the limit of large mass one is left
with
$$
{\cal F}_{\mu x , \nu y}={i\over{2\pi}}\delta(x-y)\epsilon_{\mu\nu}\eqno\enum$$
{}From ${\cal F}_{\mu x , \nu y}$ one can construct the current
$J_\mu^{\rm cov}$ to linear order in $A_\mu (x)$ and then obtain the anomaly.
The answer is just twice as large as the one obtained in eq. (3.61). This
factor indicates that now we have obtained the covariant anomaly.
Moreover, the Berry phase
computation is known to reproduce the results for other calculations
of the induced effective actions in odd-dimensional space time. These
calculations, as stressed above, are made in the
boundary free case and hence are only concerned with the bulk
effects that we already showed give the covariant rather than the consistent
anomaly.

As for the nonconservation of the charge associated with
$J_\mu^{\rm cov}$ in the global case
we note that it is not related to extra light
particles anywhere in the system; rather it has to do with some
global degrees of freedom measuring the sensitivity of the phases associated
with the two Fock vacua entering the overlap.

\section 4. Summary and outlook

In this paper, inspired by the ``wall'' realization of chiral fermions
due to Kaplan [\ref{Kaplan}], we arrived at a connection
between  the chiral fermion
determinant and the overlap of two complex vectors.
The two vectors are the ground states for the two transfer
matrices describing the propagation on the two sides of the defect in
the ``wall'' realization. This relation can be extended to the
continuum.
We analyzed the two dimensional continuum overlap in perturbation
theory. Since the ground states
have a phase ambiguity there is a corresponding ambiguity in the
imaginary part of the effective action. We can fix the
ambiguity by doing Brillouin-Wigner perturbation theory,
and disallowing choices that differ by non-local
phase redefinitions the expected form of the anomaly is obtained.
Brillouin--Wigner perturbation theory is
simply the condition that the overlap between the ground state at some
non-zero gauge field and the ground state for the free Hamiltonian is
real. This is a condition that can be easily imposed in a numerical
computation.

The above result indicates that if
the gauge fields are perturbative (i.e.
sufficiently close to zero) this condition will produce the correct anomaly
also on the lattice. An explicit check of this is still needed.
If this works out it would be
necessary to see what happens for large gauge fields.

Another phase choice,
that has the same effect in perturbation theory to the order
we went, is to use adiabatic
deformations. Here we choose a certain interpolation that connects the given
gauge configuration to zero field and demand that the change in the
eigenvectors of the Hamiltonian at any point on this interpolating
curve is orthogonal to the eigenvectors at that point. It is possible
that this choice is more attractive for numerical calculations.
In the same vein, one might consider the fact that we are only interested
in the relative phase of the two states contributing to the overlap
insufficiently represented in the above alternatives.
One may try to use adiabatic deformation of the mass term itself
from positive to negative values as a way to interlock the two phases.
This procedure will encounter obstacles when the background gauge
becomes ``singular'' for any of the intermediate mass values. It is quite
likely that these obstacles are the heart of the matter and their presence
is a good way to single out from the phase difference the part that
is physically significant.

The numerical simulations will have to evaluate the overlap for a
given background gauge configuration. The real part will be used for
the update and the imaginary part will have to be combined with the
observable being measured. All the
eigenvectors corresponding to the negative
eigenvalues of ${\rm\bf H}_\pm$ will be found by some
iterative procedure. In an updating procedure
only a single link is changed at one time and hence the matrix in
(2.14) will only change slightly; this would make it easier to find
the new eigenvectors given the old ones. It should be noted
that, to evaluate the overlap we do not
need explicitly all the negative eigenvectors. What we really need are
two bases in the spaces $Im {\cal P}_\pm$ where ${\cal P}_\pm =-{1\over 2}
\left ( {{{\bf H_\pm}}\over {\sqrt{{\bf H_\pm}^2}}} -1\right )$ are projector
operators on the filled states in the Dirac seas.

It is well known that anomalies very often can be given a topological
interpretation [\ref{Alvarg}];
 this topology has to do with smooth space or space--time and
is lost on the lattice. To make sure that the lattice has not
completely wiped out the remnants of the topological effects it
would be necessary to perform calculations on a sequence of lattices
and verify that quantities are behaving smoothly.
The nonperturbative $SU(2)$ anomaly [\ref{Witten}] will also have to be
shown to hold on the lattice.

An important point to
note is that the overlap can be zero. This will happen whenever the
number of states with eigenvalues less than one for the two transfer
matrices are different. In order to study this it would be interesting
to study the overlap near instanton configurations.

It should be clear to the reader that we do not claim to
have here a complete solution to
the problem of regularizing nonperturbatively chiral gauge theories.
Nevertheless we feel that progress has been made irrespectively of
whether chiral gauge theories will turn out to ultimately
exist or not. The flavor of the overlap formulation seems right;
for the first time we see a lattice framework that makes contact
with the important developments in our understanding of anomalies
that took place in the mid eighties and incorporates them. If one wishes
to phrase the problem
in the mnemonic advocated by Fujikawa [\ref{Fujika}] the infiniteness of the
number
of fermion fields has made the Grassmann ``measure'' ill defined and hence
not necessarily gauge invariant (despite appearances). If one likes
more the approach of  Alvarez--Gaume and Della Pietra [\ref{Etainv}] we have a
clear
separation between
the imaginary and real parts of the chiral action. The quantities we are
dealing with are
reminiscent of the $\eta$ invariant used by these authors. Instanton
effects have a clear place and the partition function can easily vanish.
Global anomalies seem also to fit in.

We would like to suggest that the right line of approach is not
to try to jump prematurely to a conclusion whether a perturbatively
consistent chiral gauge theory can exist but rather try to build
a framework which, where it to ultimately fail, would provide
a strong indication that indeed chiral theories cannot be freed
of all cutoff dependence.  Of course,
much more work will
be needed to see if the overlap is a useful approach to the problem of
non-perturbatively defining anomaly free chiral gauge theories.

\bigskip\bigskip
\centerline{\bf Acknowledgements}
\bigskip
We would like to thank T.~Banks, D.~B.~Kaplan,
M.~F.~Golterman, S.~Shenker and E.~Witten for discussions.
E.~Witten was probably aware of a connection between the chiral
determinant and some form of overlap since the mid-eighties.
This research was supported in part by the DOE under grant \#
DE-FG05-90ER40559.

\appendix A

In this appendix we provide some intermediate steps in going from
(2.6) to (2.10).

We start by noting that, if $0< m< 1$,
$\bpm$ defined in (2.7) is
strictly positive for any gauge configuration. To see this, first note
that ${\rm\bf B}^\mu$ defined by
$${\rm\bf B}^\mu_{n\alpha i,m\beta j}=
\delta_{\alpha\beta}\delta_{m,n+\hat\mu}U^{ij}_{n,\mu},\eqno\enum$$
is unitary. Therefore,
$$||{\rm\bf B}^\mu||\le 1.\eqno\enum$$
This shows that the norm of the second term in (2.7) is less than or
equal to 4. The positivity of $\bpm$ now follows.

We now define the following change of Grassmann variables to make the
first two lines in (2.6) into a simple inner product.
For $s\ge 0$, let
$$c_s=\sqrt{\bp}\rphr;\ \ c^*_s={\sqrt{\bp}}^t\lbhl;\ \
d^*_s=\sqrt{\bp}\lphl;\ \ d_s=-{\sqrt{\bp}}^t\rbhr.\eqno\enum$$
For $s< 0$, let
$$c_s=\sqrt{\bm}\rphr;\ \ c^*_s={\sqrt{\bm}}^t\lbhl;\ \
d^*_s=\sqrt{\bm}\lphl;\ \ d_s=-{\sqrt{\bm}}^t\rbhr.\eqno\enum$$
The superscript $t$ denotes the transpose operation. Starting from
(2.6) and using (A.3) and (A.4), the effective action in (2.2) becomes
$$\eqalign{e^{S_{\rm eff}(U)}=\prod_{s\ge 0}& [\det \bp]^2
\prod_{s< 0} [\det \bm]^2\cr
\int \prod_s &[dc^*_s][dc_s][dd^*_s][dd_s]
e^{-(c^*_s,c_s)-(d^*_s,d_s)}\cr
\prod_{s\le -2}&
R_-(c^*_s,d^*_s)T_-(c^*_s,d^*_s;c_{s+1},d_{s+1})R^\dagger_-(c_{s+1},d_{s+1})
\cr
&
R_-(c^*_{-1},d^*_{-1})T_0(c^*_{-1},d^*_{-1};c_0,d_0)R^\dagger_+(c_0,d_0)
\cr
\prod_{s\ge 0}&
R_+(c^*_s,d^*_s)T_+(c^*_s,d^*_s;c_{s+1},d_{s+1})R^\dagger_+(c_{s+1},d_{s+1})
\cr}\eqno\enum$$
$$R_{\pm}(c^*_s,d^*_s)=
\exp\Bigl\{(c^*_s,{1\over \sqrt{\bpm}}\bc{1\over
\sqrt{\bpm}}d^*_s)\Bigr\}\eqno\enum$$
$$R^\dagger_{\pm}(c_s,d_s)=
\exp\Bigl\{(d_s,{1\over \sqrt{\bpm}}\bcd{1\over
\sqrt{\bpm}}c_s)\Bigr\}\eqno\enum$$
$$T_{\pm}(c^*_s,d^*_s;c_{s+1},d_{s+1})=
\exp\Bigl\{(c^*_s,{1\over \bpm}c_{s+1})+(d^*_s,{1\over
{\bpm}^t}d_{s+1})
\Bigr\}\eqno\enum$$
$$T_{0}(c^*_{-1},d^*_{-1};c_0,d_0)=
\exp\Bigl\{(c^*_{-1},{1\over \sqrt{\bp}\sqrt{\bm}}c_0)
+(d^*_{-1},{1\over
{\sqrt{\bp}}^t{\sqrt{\bm}}^t}d_0)\Bigr\}\eqno\enum$$

At this point we can use the formulas in the appendix of
[\ref{Luesch}]
to convert (A.5) into operator notation. The Grassmann variables
$c^*,d^*,c,d$ go into fermion operators $\hat c^\dagger,\hat
d^\dagger,\hat c,\hat d$ respectively. These fermion operators satisfy
the following cannonical anti commutation relations:
$$\{\hat c_{n\alpha i},\hat c^\dagger_{m\beta j}\}
=\delta_{nm}\delta_{\alpha\beta}\delta_{ij};\ \ \ \ \
\{\hat d_{n\alpha i},\hat d^\dagger_{m\beta j}\}
=\delta_{nm}\delta_{\alpha\beta}\delta_{ij}.\eqno\enum$$
All other anti commutators are zero.
The operator equations
associated with (A.6)--(A.9) are
$$\hat R_{\pm}=
\exp\Bigl\{(\hat c^\dagger,
{1\over \sqrt\bpm}\bc{1\over \sqrt\bpm}\hat d^\dagger)\Bigr\}\eqno\enum$$
$$\hat R^\dagger_{\pm}=
\exp\Bigl\{(\hat d,{1\over \sqrt\bpm}\bcd{1\over \sqrt\bpm}\hat
c)\Bigr\}\eqno\enum$$
$$\hat T_{\pm}=
\exp-\Bigl\{(\hat c^\dagger,\log{\bpm}\hat c)+(\hat d^\dagger,\log
{{\bpm}^t}\hat d)
\Bigr\}\eqno\enum$$
$$\hat T_{0}=
\exp-\Bigl\{(\hat c^\dagger,{1\over 2}\log{\bp\bm}\hat c)
+(\hat d^\dagger,{1\over 2}
\log{(\bm\bp)}^t\hat d)\Bigr\}\eqno\enum$$
The operator equation for the effective action in (A.5) is
$$\eqalign{e^{S_{\rm eff}(U)}=&
\prod_{s\ge 0} [\det \bp]^2\prod_{s<0} [\det \bm]^2\cr
&<b-|\ \Bigl\{\prod_{s\le -2} [\hat R_-\hat T_-\hat R_-^\dagger]\Bigr\}
[\hat R_-\hat T_0\hat R_+^\dagger]
\Bigl\{\prod_{s\ge 0} [\hat R_+\hat T_+\hat
R_+^\dagger]\Bigr\}\ |b+>\cr}\eqno\enum$$
$<b-|$ and $|b+>$ are the boundary conditions at $s=-\infty$ and
$s=\infty$ respectively expressed as states in the operator formalism.
To obtain (2.10) all that needs to be done now is to define the fermion
operators $\hat a^\dagger$ and $\hat a$ as
$$\hat a^\dagger \equiv \pmatrix{\hat c^\dagger & \hat d\cr};\
\ \ \ \hat a \equiv \pmatrix{\hat c\cr \hat d^\dagger\cr}\eqno\enum$$
These fermion operators also satisfy canonical anti commutation
relations (equation (2.15)) and in terms of (A.16), (A.15) becomes
(2.10). A useful formula for executing this last step is
$$e^{\hat a^\dagger {\bf\rm K} \hat a}
=e^{\hat a^\dagger {\bf\rm K}_1 \hat a}
e^{\hat a^\dagger {\bf\rm K}_2 \hat a},\eqno\enum$$
where,
$$e^{\bf\rm K}=e^{{\bf\rm K}_1}e^{{\bf\rm K}_2},\eqno\enum$$
for any two matrices ${\bf\rm K}_1$ and ${\bf\rm K}_2$.

\appendix B

In this appendix we present the derivation of (2.27) from (2.25) and
(2.26).
Let us combine all the three indices $nAi$ into one big index $I$.
{}From (2.25) and (2.26), the overlap between the two ground states is
$$\eqalign{<0-|0+>=\sum_{{}^{I_1,\cdots,I_N}_{J_1,\cdots,J_N}}
&{\rm\bf L}_{1,J_1}\cdots
{\rm\bf L}_{N,J_N}{\rm\bf R}^\dagger_{I_1,1}\cdots
{\rm\bf R}^\dagger_{I_N,N}\cr
&\ \ \ <0|\hat a_{I_N}\cdots \hat a_{I_1}\hat a^\dagger_{J_1}\cdots \hat
a^\dagger_{J_N}|0>\cr}\eqno\enum$$
For a non-zero contribution to the above sum, the set
$\{J_1,\cdots,J_N\}$
has to be a permutation of $\{I_1,\cdots,I_N\}$. Further all the
$I_k$'s have to different. This yields
$$<0-|0+>=\sum_{\pi} {\rm sign}( \pi ) \sum_{I_1\ne I_2\cdots\ne I_N}
{\rm\bf L}_{\pi_1I_1}{\rm\bf R}^\dagger_{I_11}\cdots
{\rm\bf L}_{\pi_NI_N}{\rm\bf R}^\dagger_{I_NN}\eqno\enum$$
$\pi$ denotes a permutation of the set $1,\cdots,N$.
The restriction that all the $I_k$'s are different in the above sum
can be removed, because if they are included, the sum over $\pi$ will,
in any case, render then zero. The result then is (2.27).

\beginbibliography
\bibitem{Smit}
J. Smit, Nucl. Phys. B (Proc. Suppl.) 4 (1988) 451.
\bibitem{Rome}
A. Borrelli, L. Maiani, G. C. Rossi, R. Sisto and M. Testa, Nucl.
Phys. B333 (1990) 335; G. C. Rossi, R. Sarno and R. Sisto, Nucl. Phys.
B398 (1993) 101; S. A. Frolov and A. A. Slavnov, Munich preprint,
MPI-Ph 93-12, March 1993.
\bibitem{Kaplan}
D. B. Kaplan, Phys. Lett. B288 (1992) 342.
\bibitem{Narger}
R. Narayanan and H. Neuberger, Phys. Lett. B302 (1993) 62.
\bibitem{Fronov}
S. A. Frolov and A. A. Slavnov, SPhT/92-051, Saclay preprint, 1992.
\bibitem{Nieiya}
H. B. Nielsen and M. Ninomiya, Nucl. Phys. B185 (1981) 20;
Nucl. Phys. B193 (1981) 173;
Phys. Lett. 105B 219;
in ``Trieste Conference on Topological
Methods in Quantum Field Theories'', World Scientific, 1991;
D. Friedan, Comm. Math. Phys. 85 (1982) 481; F. Wilczek,
Phys. Rev. Lett. 59 (1987) 2397;
L. H. Karsten and J. Smit, Nucl. Phys. B183 (1981) 103; L. H. Karsten,
Phys. Lett. 104B (1981) 315; J. Smit, Acta Phys. Pol., B17
(1986) 531;
\bibitem{Calvey}
C. G. Callan, Jr., and J. A. Harvey, Nucl. Phys. B250 (1985) 427.
\bibitem{Kordes}
C. P. Korthals-Altes, S. Nicolis and J. Prades, CPT-93/P.2920, Center
de Physique Th\'eorique preprint, June 1993.
\bibitem{Kappri}
D. B. Kaplan, private communication.
\bibitem{Shamir}
Y. Shamir, WIS-93/20/FEB-PH, Weizmann Institute of Science preprint,
1993.
\bibitem{Disrey}
J. Distler and S-Y. Rey, PUPT-1386, Princeton University preprint, 1993.
\bibitem{Kaplat}
D. B. Kaplan, Lattice '92, Nucl. Phys. B(Proc. Suppl.)30 (1993) 597.
\bibitem{Yukawa}
D. N. Petcher, Nucl. Phys. B (Proc. Suppl.) 30 (1993) 50.
\bibitem{Luesch}
M. L\"uscher,
Commun. Math. Phys. 54 (1977) 283
\bibitem{Gonsan}
M .F .L. Golterman, K. Jansen and D. B. Kaplan, Phys. Lett. B301
(1993) 219.
\bibitem{Alvten}
L. Alvarez-Gaume and E. Witten, Nucl. Phys. B234 (1983) 269.
\bibitem{Topol}
M. G\"ockeler, A. S. Kronfeld, G. Schierholz and U. -J. Wiese, HLRZ
preprint, HLRZ 92-34, August 1992;
M. G\"ockeler and G. Schierholz, HLRZ preprint, HLRZ 92-81, November 1992.
\bibitem{Jackiw}
R. Jackiw in ``Relativity, Groups and Topology II'', Les Houches 1983,
eds. B. S. Dewitt and R. Stora, North-Holland, Amsterdam, (1984).
\bibitem{Covcon}
S. G. Naculich, Nucl. Phys. B296 (1988) 837;
W. Bardeen and B. Zumino, Nucl. Phys. B244 (1984) 421;
L. Alvarez-Gaume and P. Ginsparg, Ann. Phys. (NY) 161 (1985) 423;
\bibitem{Zumino}
B. Zumino in ``Relativity, Groups and Topology II'', Les Houches 1983,
eds. B. S. Dewitt and R. Stora, North-Holland, Amsterdam, (1984);
R. Stora in Carg\'ese lectures 1983; B. Zumino, W-Y. Shi and A. Zee,
Nucl. Phys. B239 (1984) 477.
\bibitem{Redlich}
N. Redlich, Phys. Rev. Lett 52 (1984) 18; Phys. Rev. D29 (1984) 2366;
A. J. Niemi and G. W. Semenoff, Phys. Rev. Lett 51 (1983) 2077.
\bibitem{Index}
A. J. Niemi and G. W. Semenoff, Nucl. Phys. B269 (1986) 131;
A. J. Niemi, Nucl. Phys. B253 (1985) 14;
M. Atiyah, V. Patodi and I. Singer, Bull. Math. Soc. (London) 5 (1973)
229;
Proc. Cambridge. Phil. Soc. 77 (1975) 42; 78 (1975) 405; 79 (1976) 71;
M. Ninomiya and C-I. Tan, Nucl. Phys. B257 [FS14] (85) 199.
\bibitem{Aoki}
S. Aoki and Y. Kikukawa, UTHEP-258/KUNS-1204, Univ. of Tsukuba
and Kyoto Univ. preprint, June 1993.
\bibitem{Banks}
T. Banks, Phys. Lett. 272B (1991) 75.
\bibitem{Uhol}
A. J. Niemi and G. W. Semenoff, Phys. Rev. Lett. 55 (1985) 927;
B. Simon, Phys. Rev. Lett. 51 (1983) 2167; Y-S. Wu and A. Zee, Nucl.
Phys. B258 (1985) 157.
\bibitem{Berry}
M. Berry, Proc. Royal Soc. (London) A392 (1984) 45.
\bibitem{Nienwu}
A. J. Niemi, G. W. Semenoff and Y-S. Wu, Nucl. Phys. B276 (1986) 173.
\bibitem{Alvarg}
L. Alvarez-Gaume and P. Ginsparg, Nucl. Phys. B243 (1984) 449.
\bibitem{Witten}
E. Witten, Phys. Lett. 117B (1982) 324.
\bibitem{Fujika}
K. Fujikawa, Phys. Rev. Lett. 42 (1979) 1195.
\bibitem{Etainv}
A. J. Niemi in High Energy Physics 1985, Vol I, eds. M. J. Bowick and
F. G\"ursey, World Scientific, (1986);
L. Alvarez-Gaume, Lectures delivered at the Int. School. on
Math. Phys. (Erice) July 1985 and ENS Summer Workshop, August 1985;
S. Forte, Nucl. Phys. B301 (1988) 69;
L. Alvarez-Gaume and S. Della Pietra, in ``Recent Developments in
Quantum Field Theory'',  eds. J. Ambjorn, B. J. Durhuus and
J. L. Petersen, Elsevier Science Publishers, (1985) 95;
L. Alvarez-Gaume, S. Della Pietra and V. Della Pietra,
Phys. Lett. 166B (1986) 177;
S. Della Pietra, V. Della Pietra and L. Alvarez-Gaume,
Comm. Math. Phys. 109 (1987) 691.
\endbibliography

\end